\documentclass[manuscript]{aastex}
\usepackage{emulateapj5}

\slugcomment{To appear in the Astrophysical Journal}

\shortauthors{Brunt \& Mac Low}
\shorttitle{Supersonic Turbulence}

\begin{document}

\def\gtabouteq{\,\hbox{\raise 0.5 ex \hbox{$>$}\kern-.77em
                    \lower 0.5 ex \hbox{$\sim$}$\,$}}
\def\ltabouteq{\,\hbox{\raise 0.5 ex \hbox{$<$}\kern-.77em
                     \lower 0.5 ex \hbox{$\sim$}$\,$}}
\def\vlsr{V$_{LSR}$}
\def\kms{km s$^{-1}$}

\title{Modification of Projected Velocity Power Spectra by Density Inhomogeneities in Compressible Supersonic Turbulence}

\author{Christopher M. Brunt\altaffilmark{1,2,3} \& Mordecai-Mark Mac Low\altaffilmark{4}}
\altaffiltext{1}{National Research Council, 
       Herzberg Institute of Astrophysics, 
       Dominion Radio Astrophysical Observatory, 
       P.O. Box 248, Penticton, BC, 
       V2A 6J9 Canada ; chris.brunt@nrc.ca
    }

\altaffiltext{2}{ Department of Physics and Astronomy, 
         University of Calgary, 
         2500 University Dr. NW, 
         Calgary, AB, 
         T2N 1N4 Canada
 }

\altaffiltext{3}{ Current address : Department of Astronomy, 
                  LGRT 632, University of Massachusetts, 
                  710 North Pleasant Street, Amherst, MA 01003; 
                  brunt@fcrao1.astro.umass.edu
 }

\altaffiltext{4}{ Department of Astrophysics,
                  American Museum of Natural History,
                  79th St. and Central Park West
                  New York, NY, 10024-5192, USA;
                  mordecai@amnh.org
 }

\begin{abstract}

The scaling of velocity fluctuation $\delta v$ as a function 
of spatial scale $\ell$ in molecular clouds can be measured 
from size-linewidth relations, principal component analysis, 
or line centroid variation.  Differing values of the power 
law index of the scaling relation 
$\langle \delta v^{2} \rangle^{1/2} \propto \ell^{\gamma_{3D}}$ 
in three dimensions (3D)
are given by these different methods: the first two give 
$\gamma_{3D} \simeq 0.5$, while line centroid analysis 
 gives $\gamma_{3D} \simeq 0$.
This discrepancy has previously not been fully appreciated,
as the variation of {\it projected} velocity line centroid
fluctuations ($\langle {\delta}v_{lc}^{2} \rangle^{1/2}$~$\propto$~$\ell^{\gamma_{2D}}$)
is indeed described, in 2D, by $\gamma_{2D}$~$\approx$~0.5.
However, if projection smoothing is accounted for, this
implies, in 3D, that $\gamma_{3D}$~$\approx$~0. We suggest that a
resolution of this discrepancy can be achieved by accounting for
the effect of density inhomogeneity on the observed $\gamma_{2D}$
obtained from velocity line centroid analysis.
Numerical simulations of compressible turbulence are used to
show that the effect of density inhomogeneity statistically but
not identically reverses the effect of projection smoothing
in the case of {\it driven} turbulence
so that velocity line centroid analysis does indeed predict that
$\gamma_{2D}$~$\approx$~$\gamma_{3D}$~$\approx$~0.5. 
For decaying turbulence, the
effect of density inhomogeneity on the velocity line centroids
diminishes with time such that at late times, 
$\gamma_{2D}$~$\approx$~$\gamma_{3D}$~+~0.5 due 
to projection smoothing alone. Deprojecting the
observed line centroid statistics thus requires some 
knowledge of the state of the flow. 
This information can be inferred
from the spectral slope of the column density power spectrum
and a measure of the standard deviation in column
density relative to the mean. Using our numerical results we
can restore consistency between line centroid analysis, principal
component analysis and size-linewidth relations, and we derive
$\gamma_{3D}$~$\approx$~0.5, corresponding to shock-dominated
(Burgers) turbulence, which also describes the
simulations of driven turbulence at scales where numerical
dissipation is negligible. We find that
this consistency requires
that molecular clouds are continually driven on
large scales or are only recently formed. 

\end{abstract}

\section{Introduction}

The presence of turbulence in molecular clouds
is diagnosed by the prevalence of superthermal
linewidths. A second important diagnostic is the
presence of macroscopic velocity fluctuations, and
the exponent, $\gamma_{3D}$, that describes the variation of
root mean square velocity fluctuation ($\langle {\delta}v^{2} \rangle^{1/2}$)
with spatial scale, $\ell$, in 3D : $\langle {\delta}v^{2}\rangle^{1/2}$~$\propto$~$\ell^{\gamma_{3D}}$.
The exponent, $\gamma_{3D}$, can vary depending on the
state of turbulence in the medium; e.g., the
Kolmogorov (1941) case has $\gamma_{3D}$~$\approx$~1$/$3, which
applies to incompressible turbulence and may apply
to decaying compressible turbulence (Falgarone et al. 1994);
the compressible, shock-dominated case (Burgers 1974),
has $\gamma_{3D}$~$\approx$~1$/$2. 
The scale dependence of velocity dispersion in
molecular clouds will in general be dependent on
the energy injection spectrum, in addition to
the transfer of energy between scales.
For a pure turbulent cascade of kinetic energy,
measurement of $\gamma_{3D}$ provides fundamental physical
information on the nature of interstellar turbulence. 

In this paper we describe a long-standing but largely
unappreciated discrepancy between observational
estimates of $\gamma_{3D}$. Velocity line centroid analysis,
operating in (projected) 2D suggests that $\gamma_{3D}$~$\approx$~0
when the effects of projection smoothing (von Hoerner 1951; 
O'Dell \& Castaneda 1987; Stutzki et al. 1998; Miville-Desch{\^e}nes,
Levrier, \& Falgarone 2003; Brunt et al. 2003) are accounted
for. Size-linewidth relations (e.g. Solomon et al. 1987) and 
principal component analysis (Heyer \& Schloerb 1997; Brunt et al. 2003)
suggest that  $\gamma_{3D}$~$\approx$~0.5.

We argue in this paper that, for driven turbulence, density inhomogeneity
induces small scale structure on the projected
velocity line centroid field such that the
effects of projection smoothing are statistically
but not identically reversed and thus that 
the observed $\gamma_{2D}$
obtained from velocity line centroid analysis 
does in fact closely approximate the true
3D index, $\gamma_{3D}$. Some indication
that this is true was found by Ossenkopf \& Mac Low (2002).

We begin with an overview of the 
statistical properties of velocity fields in
3D, and in projected 2D as seen by velocity line
centroid analysis. In {\S}~\ref{sec:numerical} we
describe our numerical data, and in 
{\S}~\ref{sec:observational} we describe our
observational data. The results of velocity
line centroid analysis applied to the numerical
simulations of turbulence, both with and without
density weighting, are discussed in {\S}~\ref{sec:numericalresults}.
In {\S}~\ref{sec:observationalresults} we
estimate the observed power spectrum of line centroid
velocity fluctuations in the NGC~7129 molecular cloud,
and interpret the measured power spectrum exponent and
previously measured line centroid structure function exponents, 
accounting both for projection smoothing and density
inhomogeneity.

\section{Velocity Line Centroid Analysis : Theory{\label{sec:linecentroid}}}

There have been a number of correlation studies on
projected line centroid velocities in molecular 
clouds (Kleiner \& Dickman 1985;
Kitamura et al. 1987; Hobson 1992; Miesch \& Bally 1994; 
Ossenkopf \& Mac Low 2002).
Most of the studies have proceeded in direct space, relying in
large part on the formalism established by Kleiner \& Dickman (1985)
and Dickman \& Kleiner (1985). Here we use the related Fourier space
representation, and measure line centroid velocity fluctuations
via their power spectrum. The power spectrum ($P(k)$ as
a function of wavenumber $k$) of velocity fluctuations is:
$P(k)$~$\propto$~$k^{-\kappa_{ND}}$, where $\kappa_{ND}$~=~$2\gamma_{ND} + N$
in $N$ dimensions\footnote{Note that for $\kappa_{ND}$~$<$~$N$, $\gamma_{ND}$ saturates to $\sim$~zero, while for $\kappa_{ND}$~$>$~$N+2$, $\gamma_{ND}$ saturates to unity -- see Stutzki et al. (1998) and see Brunt et al. (2003) for a demonstration of the latter case.}. The above relationship between
$\kappa_{ND}$ and $\gamma_{ND}$ applies to the case of exact
power law power spectra. In general, the power spectrum and
structure function of fluctuations are {\it integrally} to
each other (the structure function is a simple transformation
of the autocorrelation function which in turn is the
inverse Fourier transform of the power spectrum) and defining
the statistical relationship between them with single indices is difficult.
Our observational data does have well-defined power law
behavior over the observed range of scales, so the above
simple prescription is valid.

We use the power spectrum for three reasons: (1) in the numerical
fields, increasing separations ultimately decrease due to
field periodicity resulting in artificially smaller $\gamma_{ND}$
for a given $\kappa_{ND}$ (see Brunt et al. 2003); (2) the
numerical fields have power spectra that are not power laws
over all wavenumbers due to the scales at which the
turbulence was driven and the effects of numerical dissipation--
we use a Fourier space scheme that avoids problems arising
from this; (3) for the observational data, corrections for
instrumental noise and beam-smearing are easily achieved
in Fourier space--any direct space beam-smearing corrections must in
general proceed via Fourier space anyway. 
We thus characterize all measurements
in terms of power spectra, but note the direct space
equivalent according to the simple $\kappa_{ND}$--$\gamma_{ND}$
given above.

In the power spectrum representation,
the Kolmogorov case has $\kappa_{3D}$~=~11$/$3
and the Burgers case has $\kappa_{3D}$~=~4.
Previously, the power spectrum approach
has been used by Kitamura et al. (1987), and, more recently,
by Miville-Desch{\^e}nes et al. (2003(b)) for \ion{H}{1}~21~cm line
emission.

The true projected velocity line centroid field is :
$$ v_{lc}^{true}(x,y) \;\;\; = \;\;\; \sum_{i=1}^{N_{pix}} \; v_{z}(x,y,z_{i}) \;\; / \; N_{pix} ,\;\; \eqno(1)$$
where $v_{z}(x,y,z)$ is the line of sight velocity component and $N_{pix}$ is the number of 
spatial pixels along each line of sight (this is replaced by a similar integral form for
the continuous case; our numerical and observational data are pixelized, so we
will use the pixelized equations here). 

Consider first an unweighted projection of the
velocity field over the line of sight ($z$) axis.
Generation of a velocity centroid map, by projection over
the $z$ direction, simply selects the
wavevectors for which k$_{z}$~=~0, since all
others are oscillatory along the $z$ direction
and will average out (disregarding inhomogeneous
sampling). 
For an isotropic power law power spectrum in 3D, the projected 
velocity line centroid 2D power
spectrum has the same spectral slope as it does in 3D; i.e.
$\kappa_{2D}$~=~$\kappa_{3D}$ (see e.g. Stutzki et al. 1998;
Miville-Desch{\^e}nes et al. 2003; Brunt et al. 2003); more
precisely, the observed (projected) 2D power spectrum is 
identical to the k$_{z}$~=~0 plane in the 3D power spectrum.
Only the transverse component of the velocity field
is projected into 2D (e.g. Kitamura et al. 1987); in other
words : the longitudinal contribution to $v_{z}$
has no amplitude on the k$_{z}$~=~0 plane.

For the case of a power law power spectrum, while
the spectral slope is unchanged on going from 3D
to projected 2D, the direct space statistical properties of
the field {\it are} modified. The variation of
root mean square velocity fluctuation ($\langle \delta v^{2} \rangle^{1/2}$) with
spatial scale ($\ell$) is : 
$$ \langle \delta v^{2} \rangle^{1/2} \;\;\; \propto \;\;\; \ell^{(\kappa_{ND} - N)/2} \;\;\; \propto \;\;\; \ell^{\gamma_{ND}} ,\;\; \eqno(2)$$
where $N$ is the dimension on which the field is defined (here, $N$~=~2~or~3)
and where $v$, for the 2D case, is understood to be the projected line centroid field, $v_{lc}^{true}$.
The criterion $\kappa_{3D}$~=~$\kappa_{2D}$ leads to $\gamma_{2D}$~=~$\gamma_{3D}$~+~0.5;
this is projection smoothing. For the Kolmogorov (1941)
case ($\kappa_{3D}$~=~$\kappa_{2D}$~=~11/3 ; $\gamma_{3D}$~=~1/3) the 
projected index is $\gamma_{2D}$~=~5/6 (von Hoerner 1951; see O'Dell \& Castaneda 1987).
For Burgers (1974) compressible turbulence, we have $\kappa_{3D}$~=~$\kappa_{2D}$~=~4,
$\gamma_{3D}$~=~0.5, and $\gamma_{2D}$~=~1 due to projection smoothing. 

An observed
$\gamma_{2D}$~=~0.5 thus implies $\gamma_{3D}$~=~0. Only in the case that
the line of sight depth over which the observational tracer exists is much less than the
total transverse scale of emission observed on the sky (Miville-Desch{\^e}nes et al. 2003;
see also O'Dell \& Castaneda 1987) is it true that $\gamma_{2D}$~$\approx$~$\gamma_{3D}$;
i.e. this requires thin, sheet-like clouds, oriented face-on towards us.

The density weighted
projected velocity line centroid field is :
$$ v_{lc}^{wgt}(x,y) \;\;\; = \;\;\;  \sum_{i=1}^{N_{pix}} n(x,y,z_{i}) \; v_{z}(x,y,z_{i}) \;\; / \; \sum_{i=1}^{N_{pix}} \; n(x,y,z_{i}) , \;\; \eqno(3)$$
where $n(x,y,z)$ is the density field. For the case of uniformly excited,
optically thin gas, Equation~3 is the appropriate form for predicting
the velocity line centroid field. As recognized by Kleiner \& Dickman (1985),
the form of Equation~3 suggests that the projected ``velocity line centroid''
field is more akin to a measure of the line-of-sight momentum of the
medium; however, the normalization by the projected column density
complicates this assignment (dimensionally, the line centroid field has
units of velocity). For uniform density (e.g. the case of
fully incompressible turbulence) the density field amplitude is
normalized out of the line centroid field. More generally, for
inhomogeneous density, the mean projected density field is normalized out,
but variations in density along the line of sight may induce additional
structure in the measured velocity line centroid field.

Previously established results of
relevance here are : (1) Miville-Desch{\^e}nes et al. (2003)
showed that density inhomogeneity does
not affect velocity line centroids in the case of uncorrelated
density and velocity fields (thus preserving $\gamma_{2D}$~=~$\gamma_{3D}$~+~0.5).
However Miville-Desch{\^e}nes et al. (2003) used stochastic
density and velocity fields (including some degree of non-physical
correlation between density and velocity).
(2) Brunt et al. (2003)
found that density inhomogeneity {\it does} induce additional
small-scale structure on the velocity line centroid field
in the case of physically correlated density and velocity (resulting
in $\gamma_{2D}$~$<$~$\gamma_{3D}$~+~0.5). 
They also found that the correlations between 
density and velocity are important (for a given amplitude of
density fluctuations). Brunt et al. (2003)
used ``spectrally-modified'' numerically-simulated
density and velocity fields (see Lazarian et al. 2001)
in which the density-velocity coupling is also modified to some extent. 
(3)  Ossenkopf \& Mac Low (2002) examined
numerical simulations of turbulence, and measured
$\gamma_{3D}$ in 3D and $\gamma_{2D}$ from the
projected line centroid velocity field of the
simulations (including density-weighting). They found that
$\gamma_{3D}$~$\approx$~$\gamma_{2D}$~$\approx$~0.5;
they also noted that {\it projection smoothing does not
predict this behavior}.
This is an indication that the effect of density weighting
counters the effects of projection smoothing.
(4) Lazarian \& Esquivel (2003, submitted) have inferred that
density inhomogeneity does {\it not} affect the power spectrum
of velocity line centroid fluctuations.

Observationally, the density weighted line centroid velocity field
may be approximated by use of an optically thin tracer :
$$ v_{lc}^{obs}(x,y) \;\;\; = \;\;\;  \sum_{i=1}^{N_{chan}} \; T(x,y,v_{z_{i}}) \; v_{z_{i}}(x,y) \;\; / \; \sum_{i=1}^{N_{chan}} \; T(x,y,v_{z_{i}}) \;\; \eqno(4) ,$$
where $T(x,y,v_{z_{i}})$ is the observed radiation temperature,
$v_{z_{i}}(x,y)$ is the velocity of the $i^{th}$ channel
at position $x,y$, and $N_{chan}$ is the number of channels along each
line of sight.

\section{Numerical Data~{\label{sec:numerical}}}

We use simulations of randomly driven hydrodynamical (HD) turbulence 
and magnetohydrodynamical (MHD) turbulence (Mac Low 1999), and 
decaying hydrodynamical turbulence (Mac Low et al. 1998). These
simulations were performed with the astrophysical MHD code
ZEUS-3D\footnote{Available from the Laboratory for
Computational Astrophysics at http://cosmos.ucsd.edu/lca/} (Clarke
1994), a 3D version of the code described by 
Stone \& Norman (1992a, b) using second-order advection (van Leer 1977), 
that evolves magnetic fields using constrained transport (Evans \& 
Hawley 1988), modified by upwinding along shear Alfv{\'e}n characteristics
(Hawley \& Stone 1995). The code uses a von Neumann artificial
viscosity to spread shocks out to thicknesses of three or four zones
in order to prevent numerical instability, but contains no other
explicit dissipation or resistivity. Structures with sizes close to
the grid resolution are subject to the usual numerical dissipation,
however.

The simulations were performed on a 3D,
uniform, Cartesian grid with periodic boundary conditions 
in every direction. All length scales are in fractions of the cube
size; i.e. all wavevectors run from $k$~=~1 to $k$~=~$N_{pix}$/2 for a
simulation of $N_{pix}$ pixels in each dimension.
An isothermal equation of state was used in the computations,
with sound speed chosen to be $c_s = 0.1$ in arbitrary units.  The
initial density and, where applicable, magnetic field were both
initialized uniformly on the grid, with the initial density $\rho_0 =
1$ and the initial field parallel to the $z$-axis.

The turbulent flow is initialized with velocity perturbations drawn
from a Gaussian random field determined by its power distribution in
Fourier space, following the usual procedure. As discussed in detail
in Mac Low et al.\ (1998), the decaying
turbulence runs were initialized with a flat spectrum 
with power from $k_d = 1$ to $k_d = 8$ because that will 
decay quickly to a turbulent state.  Note that
the dimensionless wavenumber $k_d$ counts the number of
driving wavelengths $\lambda_d$ in the box.  A fixed pattern of
Gaussian fluctuations drawn from a field with power only in a narrow
band of wavenumbers around some value $k_d$ offers a very simple
approximation to driving by mechanisms that act on that scale. For
driven turbulence, such a fixed pattern was normalized to produce a
set of perturbations $\delta \nu (x,y,z)$. At every time step
a velocity field $\delta\vec{v}(x,y,z) = A \delta\vec{\nu}$ was added to the
velocity $\vec{v}$, with the amplitude $A$ now chosen to maintain
constant kinetic energy input rate, as described by Mac Low (1999).

A summary of the models is given in Table~\ref{tab:numericaldata}.

\section{Observational Data~{\label{sec:observational}}}

Our observational data are $^{12}$CO~(J=1--0) and
$^{13}$CO~(J=1--0) spectral line imaging observations
acquired with the Five College Radio Astronomy Observatory
(FCRAO) 14~m telescope during the Spring season, 2003. These data
form part of a large ($\sim$~200~square degree) molecular 
line survey (in progress) within the ongoing multiwavelength Canadian 
Galactic Plane Survey (CGPS; Taylor et al. 2003).

The FCRAO 14~m telescope was equipped with the
SEQUOIA focal plane array (Erickson et al. 1999), which was
recently expanded to 32~pixels. The array is rectangular
(4~$\times$~4 pixels) with two sets of 16 pixels in different
polarizations having the same pointing centers.
Due to the expanded physical
size of the array in the dewar, it is no longer possible
for the dewar to rotate and so maintain a constant orientation
with the celestial coordinate system (our mapping grid here
is in Galactic $l,b$ coordinates). To allow mapping in
celestial coordinates, FCRAO has developed an on-the-fly (OTF)
mapping system in which the telescope is driven across the
map area and spectra are dumped on the desired grid. One
pixel near the center of the array (central pixel hereafter)
is selected to define the origin of the map coordinates. 
The other pixels traverse
the map along paths determined by the pixel separation
in the array and the current orientation
of the celestial coordinate system to the telescopic 
(azimuth-elevation; az-el) system. If the central pixel is
scanned so that it Nyquist samples the map area, then
the combination of data from all the other pixels
results in a highly (but irregularly) oversampled  map. 
In order to map large areas,
the sampling rate of the central pixel can be relaxed 
slightly. Our data are obtained by dumping the correlators
every 22.5\arcsec~(1/2 beamwidth at the $^{12}$CO frequency)
along the scan direction during a 
total scan length (in $l$) of 0.5$^{\circ}$. The 
$b$ coordinate of the central pixel is then incremented
by 33.75\arcsec~and the next scan in $l$ is taken.
The pixel separation in the array is 88.56\arcsec.
In the case that the $l,b$ system is aligned 
with the az-el system, then after two scans, the
first row of the array is located at 66.45\arcsec~in $b$ above
its original position, or 22.11\arcsec~below the
original $b$ coordinate of the second row. The next scan
places the first row of the array 11.64\arcsec~above the
first scan of the second row of the array and 12.11\arcsec~below 
the second scan of the second row. For subsequent scans,
a similar progression of sampling ensues. Apart from at the
extreme map edges, a sampling finer than the Nyquist rate is
achieved, but the sampling is slightly irregular.
More generally, the $l,b$ system is rotated with respect to the az-el
system and an even finer sampling is achieved.
An OFF (reference) measurement is taken between
every two scans of the array 
(i.e. OFF, scan out and back, OFF, scan out and back, {\it etc}),
and a calibration measurement is taken after every 4 ``out and back''
sequences ($\approx$ every 9-10 minutes).
The data are calibrated to the $T_{A}^{*}$ scale
using the standard chopper
wheel method (Kutner \& Ulich 1981).
Following recording of the data, the spectra are 
converted onto a regular grid of 22.5\arcsec~pixel
spacing using the FCRAO {\it otftool} software.

The $^{12}$CO and $^{13}$CO spectral lines were acquired
simultaneously using the recently added dual channel
correlators, which provide a total bandwidth of 50~MHz,
with 1024~channels at each frequency. The central velocity
of the spectra is $v_{lsr}$~=~--45~km~s$^{-1}$, resulting in a total
velocity coverage of $v_{lsr}$~$\approx$~--110~km~s$^{-1}$ to
$v_{lsr}$~$\approx$~+20~km~s$^{-1}$ with a channel spacing of
$\sim$~49~kHz ($\approx$~0.13~km~s$^{-1}$). The spatial
resolution of the data (beam FWHM) is 45\arcsec~for 
$^{12}$CO and 46\arcsec~for $^{13}$CO.

The {\it otftool} software was configured to remove
first order baselines derived from the spectra
in the velocity intervals: --110 to --95~km~s$^{-1}$ and
+7 to +20~km~s$^{-1}$. Spectra from individual pixels in
the array contributing to a single position in the
output grid were weighting by the reciprocal of their
noise variance derived from the above velocity intervals.
Following initial gridding, we removed additional 
($\approx$ sinusoidal) baseline
structure arising from a standing wave between the subreflector
and dewar using Fourier 
methods. First, we defined the
``spectral window'' between --20~km~s$^{-1}$ and 
+5~km~s$^{-1}$ which contains the detectable emission
in the region of interest here. Then
we interpolated over this window using a first order
baseline derived from the adjacent 100 channels on
either side of the spectral window. The emission (+ noise)
in the spectral window at each spatial position was replaced 
by the interpolated line at that position and the
resulting spectrum was converted to Fourier space using
a Fast Fourier Transform (FFT). The
first two wavevectors, of wavelength 50~MHz and 25~MHz 
($\approx$~130 and 65~km~s$^{-1}$ respectively) were
extracted from the FFT'ed spectrum, converted back to
direct space and then subtracted from the original spectrum.

The resulting sensitivities (1$\sigma$, $T_{A}^{*}$) of
the processed data are $\sim$~0.5~K and $\sim$~0.25~K 
per channel for $^{12}$CO and $^{13}$CO respectively.

\section{Velocity Line Centroid Analysis : Numerical Results{\label{sec:numericalresults}}}

\subsection{Overview~{\label{sec:dkoverview}}}

The numerical fields may be used 
to evaluate the role of density weighting 
on the power spectrum of the line centroid velocity
field. We calculated $v_{lc}^{true}$ via Equation~1
and $v_{lc}^{wgt}$ via Equation~3 for all simulations.
For the MHD models we examined line centroid fields projected
over two axes: parallel and perpendicular to the mean magnetic
field direction.
As the simulated velocity fields have periodic boundary
conditions, the $k_{z}$~=~0 plane in the 3D power spectrum
of $v_{z}$ is identical to the 2D power spectrum for $v_{lc}^{true}$.

\begin{figure}
\plotone{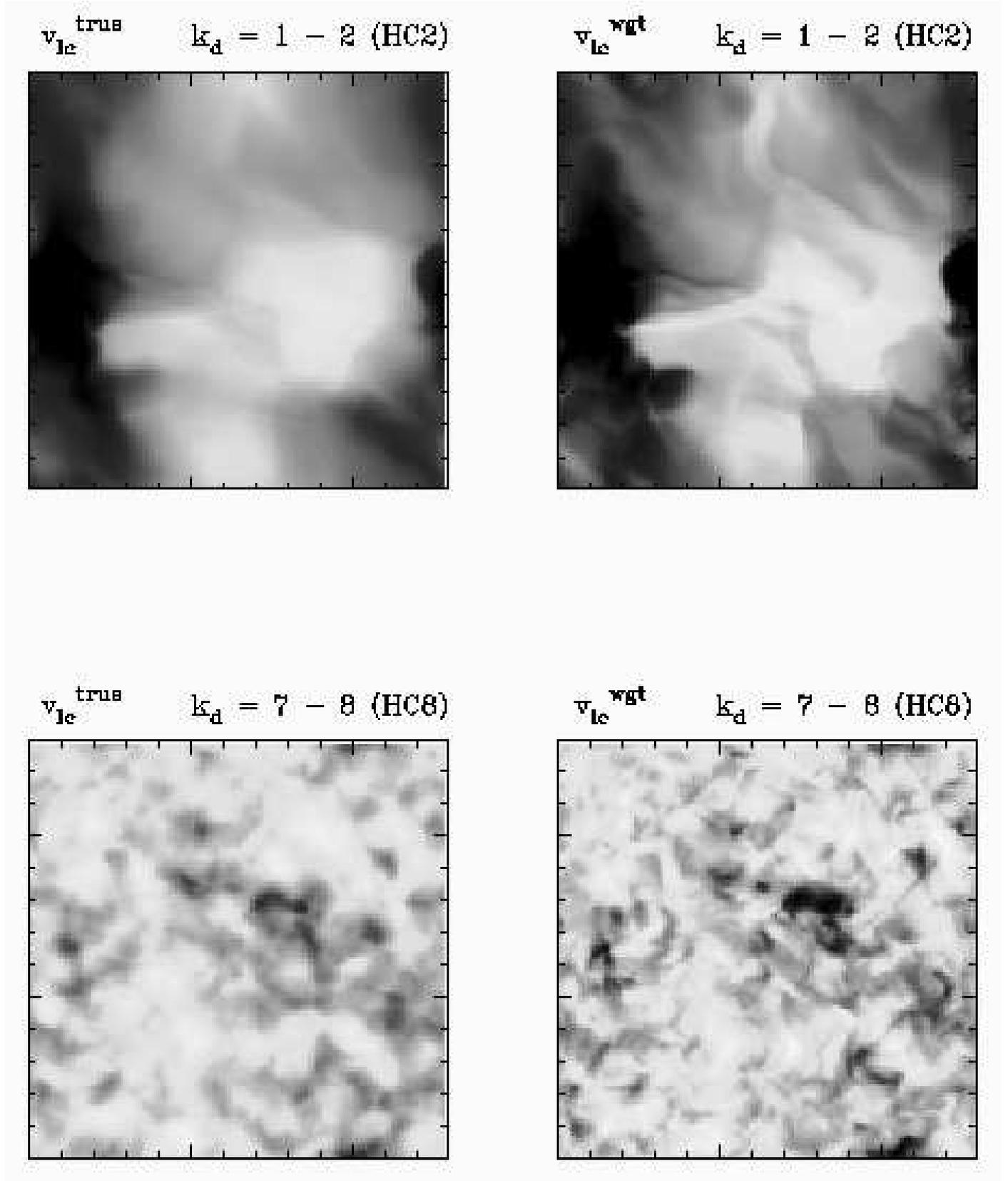}
\caption{Example velocity line centroid fields obtained from the numerical simulations, for the unweighted case ($v_{lc}^{true}$) and the density weighted case ($v_{lc}^{wgt}$).}
\label{fig:numlc}
\end{figure}

First, we examine the driven simulations.
Example velocity line centroid fields are
shown in Figure~\ref{fig:numlc}. It is immediately evident that
the inclusion of density-weighting results in 
the insertion of additional small-scale structure
into the line centroid fields.

\begin{figure}
\plotone{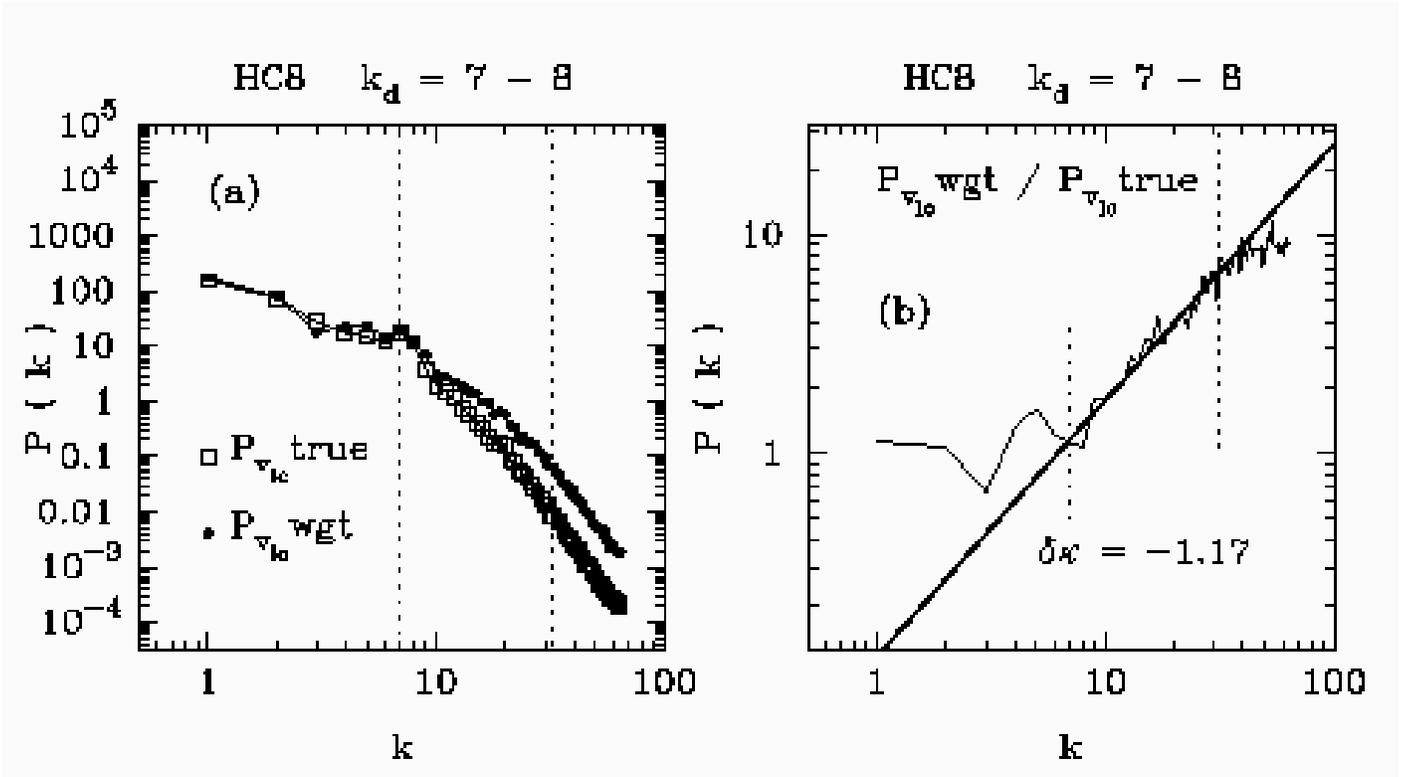}
\caption{(a) Example velocity line centroid power spectra derived from the numerical simulations (HC8 ; $k_{d}$~=~7--8), for the unweighted case ($P_{v_{lc}^{true}}$) and the density weighted case ($P_{v_{lc}^{wgt}}$). (b) The effect of density weighting determined by the spectral slope, $\delta\kappa$, of $P_{v_{lc}^{wgt}}$$/$$P_{v_{lc}^{true}}$. The dashed lines mark the range of $k$ used for the fit (heavy line) to derive $\delta\kappa$ : from $k$~=~7 (the largest driving scale) to $k$~=~32.}
\label{fig:examplefig}
\end{figure}

To quantify the effect seen in Figure~\ref{fig:numlc}, we calculated
the power spectra ($P(k)$ as function of wavenumber $k$)
of all the line centroid fields,
using a FFT. Figure~\ref{fig:examplefig}(a) shows 
representative results for HC8 model
($k_{d}$~=~7--8). There is a turnover in the power spectra
at or near $k_{d}$. For $k$~$<$~$k_{d}$, the true and weighted
line centroid power spectra are quite similar, while 
for $k$~$>$~$k_{d}$ the power spectrum of
$v_{lc}^{wgt}$ is shallower.
We can quantify the effect of density weighting
quite precisely by examining the difference (in log space) 
between the power spectra of $v_{lc}^{wgt}$ and $v_{lc}^{true}$
(in real space this corresponds to measuring the spectral index of
the ratio $P_{v_{lc}^{wgt}}$/$P_{v_{lc}^{true}}$). We characterize
the ratio of power spectra by a scaling exponent $\delta\kappa$,
where : 
$P_{v_{lc}^{wgt}}$/$P_{v_{lc}^{true}}$~$\propto$~$k^{-\delta\kappa}$
(i.e $log(P_{v_{lc}^{wgt}})$--$log(P_{v_{lc}^{true}})$~=~$const$~+~$\delta\kappa$~$log(k)$). This procedure avoids potential
problems arising from non-power-law projected
velocity spectra.

The ratio $P_{v_{lc}^{wgt}}$/$P_{v_{lc}^{true}}$ for the HC8 model
is shown in Figure~\ref{fig:examplefig}(b).
There is little modification to the true line centroid power
spectrum for spatial scales above the driving scale ($k$~=~7--8);
$P_{v_{lc}^{wgt}}$/$P_{v_{lc}^{true}}$ does not vary much with $k$ at small $k$.
This is because there are no major density enhancements on these large
scales (Ballesteros-Paredes \& Mac Low 2002). For $k$~$>$~$k_{d}$,
$P_{v_{lc}^{wgt}}$/$P_{v_{lc}^{true}}$ is a well developed
power law, with some deviation at very high $k$. We fitted the
exponent $\delta\kappa$~=~--1.17~$\pm$~0.06 over the 
range 7~$\leq$~$k$~$\leq$~32 (i.e. from
the largest driving scale to a wavelength of 4~pixels). 
The largest scale at which dissipation effects are first seen 
in these simulations is $\sim$10 zones or $k$~$\approx$~12.8 
(Ossenkopf \& Mac Low 2002) 
but we do not see any spectral breaks in the power spectrum ratio 
before $k$~=~32; we will examine the effects of this in greater detail
in Section~\ref{sec:dk12}.

\begin{figure}
\plotone{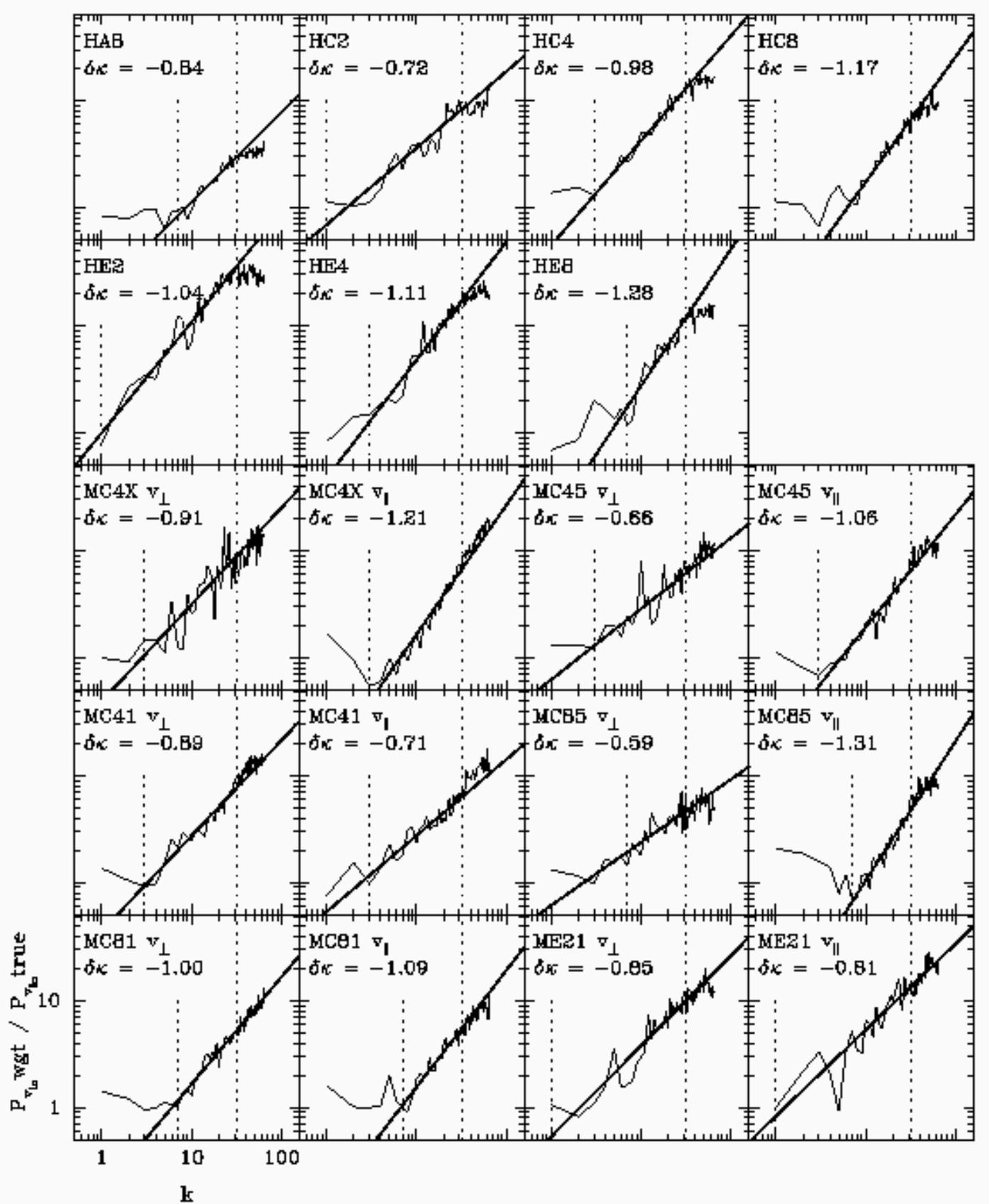}
\caption{Variation of $P_{v_{lc}^{wgt}}$$/$$P_{v_{lc}^{true}}$ with $k$ for the HD and MHD fields. Fits (heavy lines) to derive $\delta\kappa$ were made in the range of $k$ delimited by the dashed lines (largest driving scale to $k$~=~32).}
\label{fig:espechd}
\end{figure}

The other driven models gave similar results. 
Figure~\ref{fig:espechd}
shows $P_{v_{lc}^{wgt}}$/$P_{v_{lc}^{true}}$
for all driven models. These spectra were fitted between the largest
driving scale and $k$~=~32 for all models to derive $\delta\kappa$,
which in all cases was negative and $\sim$~--1 with variations
of up to $\sim$~$\pm$~0.4. The measured $\delta\kappa$ are
listed in Table~\ref{tab:numericaldata}. There is no obvious
systematic difference between the HD and MHD cases, nor
any obvious variation with rms Mach number, $M$, for the
driven runs. There is some evidence that
$\delta\kappa$ is more negative when the line of sight
is parallel to the mean magnetic field direction, but
this does not universally apply. 
Taking the observed range of $\delta\kappa$ as a natural
variation, we derive a mean (unweighted)
$\delta\kappa$~=~--0.96~$\pm$~0.22 from all the driven models.
In other words, inclusion of density weighting results
in a shallower spectral slope, with index $\sim$1 less than
the true line centroid power spectrum index. In direct
space, such a  modification would thus correspond to
$\gamma_{2D}$~$\approx$~$\gamma_{3D}$ rather than
$\gamma_{2D}$~=~$\gamma_{3D}$~+~0.5. (In general, 
$\gamma_{2D}$~=~$\gamma_{3D}$~+~0.5~+$\delta\kappa$/2.)

If the driven turbulence conditions are representative of
the conditions within molecular clouds, the scaling
exponents measured from line centroid structure functions are
approximately equal to the true, 3D exponents. 
We stress that the ``cancellation'' of projection
smoothing by small scale line centroid structure induced by density 
inhomogeneity is statistical
in nature, and the effect of density weighting does not of course
{\it identically} reverse the effects of projection smoothing.

\begin{figure}
\plottwo{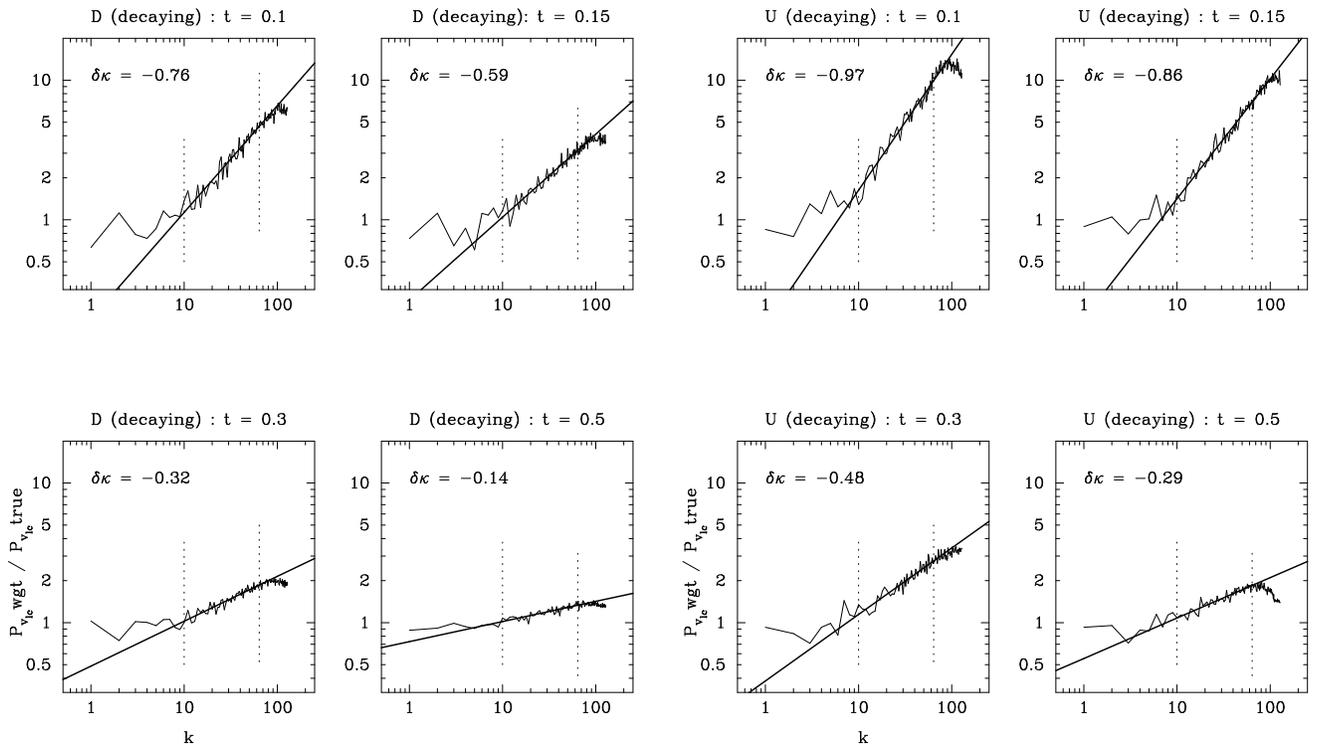}{f4b.eps}
\caption{Variation of $P_{v_{lc}^{wgt}}$$/$$P_{v_{lc}^{true}}$ with $k$ for the decaying turbulence simulations as time progresses. Fits (heavy lines) to derive $\delta\kappa$ were made in the range of $k$ delimited by the dashed lines ($k$~=~10 to $k$~=~64). Each plot is labeled by the time step in units of the sound crossing time. (Left: model D, initial $M$~=~5. Right: model U, initial $M$~=~50.)}
\label{fig:decay}
\end{figure}

\begin{figure}
\plotone{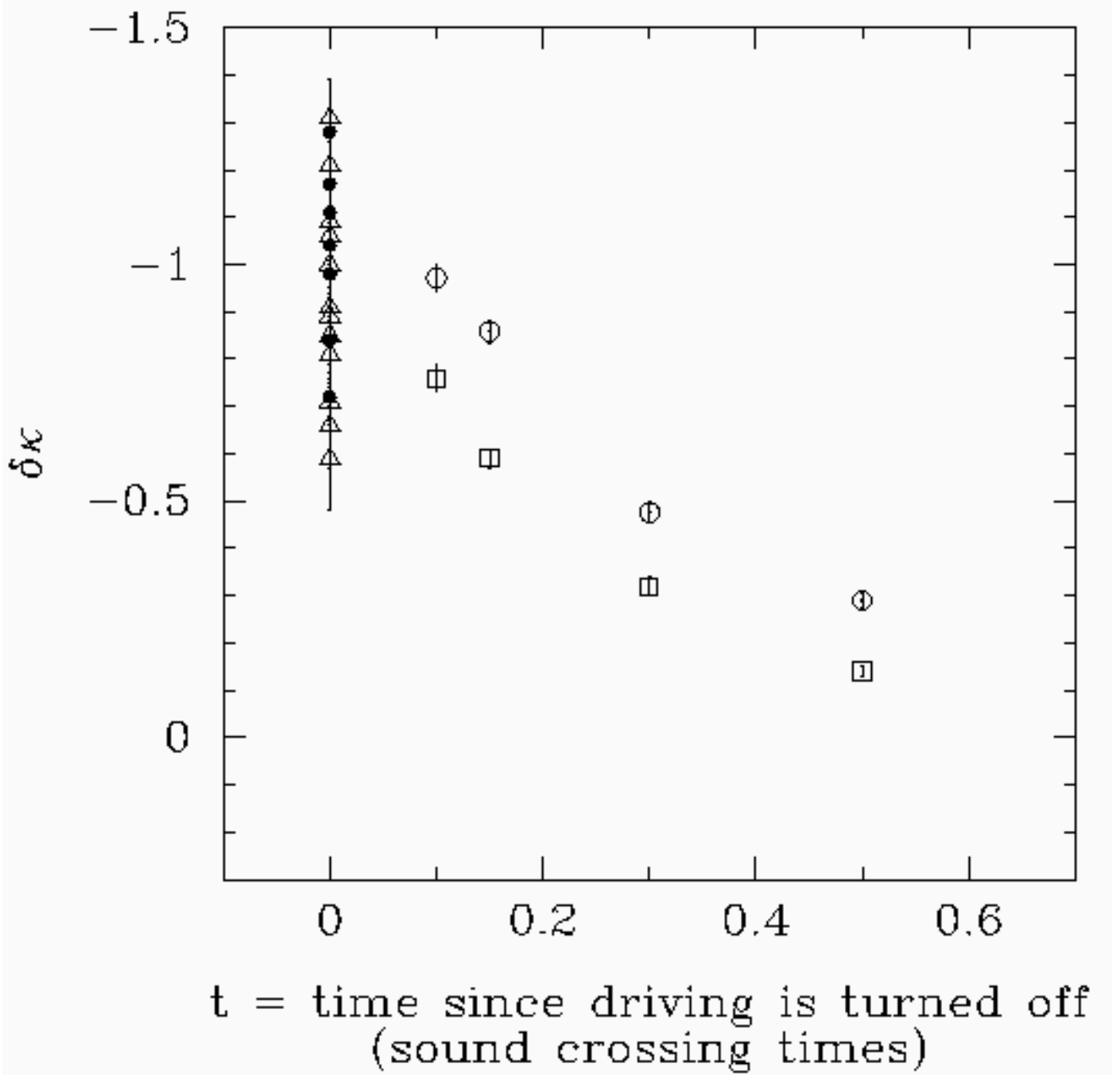}
\caption{Plot of $\delta\kappa$ versus $t$ = time after driving is turned off in units of the sound crossing time. For the driven runs, we have taken t~=~0. Filled circles are driven HD; open triangles are driven MHD; open squares are decaying HD (model D, initial $M$~=~5); open circles are decaying HD (model U, initial $M$~=~50).}
\label{fig:dtime}
\end{figure}

The same procedure to derive $\delta\kappa$ was then applied 
to the simulations of decaying turbulence. Figure~\ref{fig:decay}
shows the plots of $P_{v_{lc}^{wgt}}$/$P_{v_{lc}^{true}}$ evolving
as time increases. We have fitted
$\delta\kappa$ over the range 10~$\leq$~$k$~$\leq$~64 for all fields
(motivated again by the lack of clear breaks in the power
spectrum ratio)
and these are listed in Table~\ref{tab:numericaldata}.
This shows that as the turbulence decays, the effect of density
inhomogeneity on the line centroid field is reduced, due to
the lessened density contrast. If molecular clouds
are in such a state, then projection smoothing {\it only} applies
to the observed line centroid field, and 
$\gamma_{2D}$~$\approx$~$\gamma_{3D}$~+~0.5.
Over the course of one sound crossing time, the density
contrast is reduced sufficiently, even for initially highly supersonic
turbulence, that $\delta\kappa$ becomes
very small, due to weaker shocks in decaying turbulence
relative to those in driven turbulence
(Smith, Mac Low, \& Zuev 2000; Smith, Mac Low, \& Heitsch 2000); in other
words, if the turbulence is continually driven, or has only recently
entered a decaying state, then we expect $\delta\kappa$~$\approx$~--1,
but this evolves quickly to $\delta\kappa$~$\longrightarrow$~0 over a sound
crossing time. Figure~\ref{fig:dtime} summarizes the evolution
of $\delta\kappa$ after the driving is turned off.

\subsection{Regime of Validity~{\label{sec:dk12}}}

\begin{figure}
\plotone{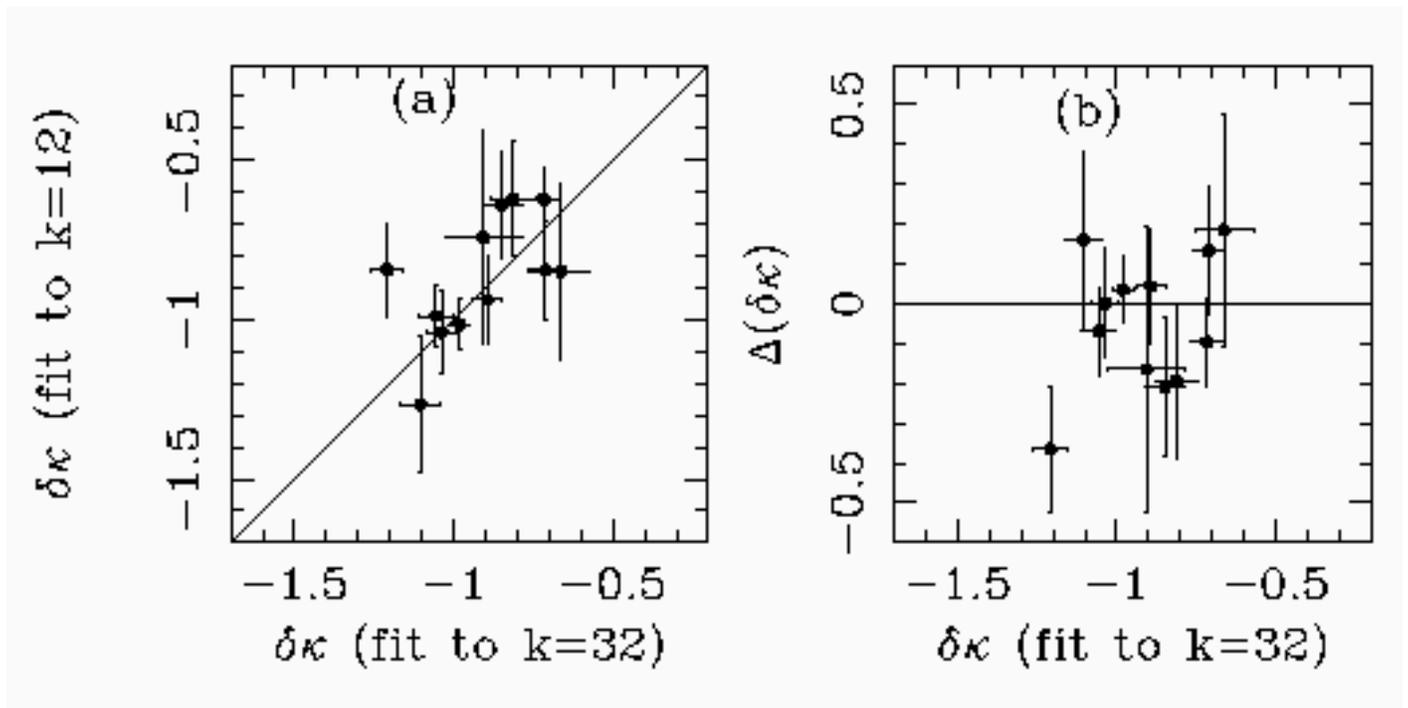}
\caption{a) Plot of $\delta\kappa$ derived from fits to the power spectrum ratio fitted up to $k$~=~12 versus $\delta\kappa$ derived from fits to the power spectrum ratio fitted up to $k$~=~32, for a subset of the driven runs (see text). The solid line is a line of equality.  (b) Difference between the $\delta\kappa$ values shown in (fit to $k$~=~32 minus fit to $k$~=~12). The solid line marks a difference of zero.}
\label{fig:ddk}
\end{figure}

As noted above, we are fitting $\delta\kappa$ over a
range of $k$ that includes the dissipation region
($k$ greater than $\sim$~12.8), motivated by the
lack of distinct breaks in the power spectrum ratio.
To test whether dissipative effects could induce
systematic biases in our results, we repeated the
fits while restricting the range to a maximum $k$~=~12.
This was done only for the driven models for which 
$k_d$~=~1--2 and $k_d$~=~3--4 
in order that sufficient data was still available
for the fits. The resulting measurements of
$\delta\kappa$ are plotted against the
measurements of $\delta\kappa$ fitted to
$k$~=~32 in Figure~\ref{fig:ddk}. For this
sample, we find that the mean $\delta\kappa$
are --0.91~$\pm$~0.17 (fitted to $k$~=~32)
and --0.87~$\pm$~0.19 (fitted to $k$~=~12).
These are not significantly different,
so our visual conclusion that
there are no breaks in the power spectrum 
ratios is good.

The advantage of using a relative measure, $\delta\kappa$,
in the above analysis is that it circumvents 
potential problems arising from small curvatures
in the intrinsic velocity power spectra. We now
examine actual measurements of the spectral indices
of the projected velocity power spectra. These
absolute indices are needed to examine
the range of validity of $\delta\kappa$ as derived
above. Due to the curvature in the
power spectra, likely arising from dissipative
effects at higher $k$, the absolute spectral
slopes depend on the range in $k$ over which
they are derived. In what follows, we label
the spectral slope of the unweighted projected
velocity field as $\kappa_{2D,u}$ and the spectral
slope of the density-weighted projected
velocity field as $\kappa_{2D,w}$. These are
related by $\kappa_{2D,w}$~--$\kappa_{2D,u}$~=~$\delta\kappa$.

\begin{figure}
\plotone{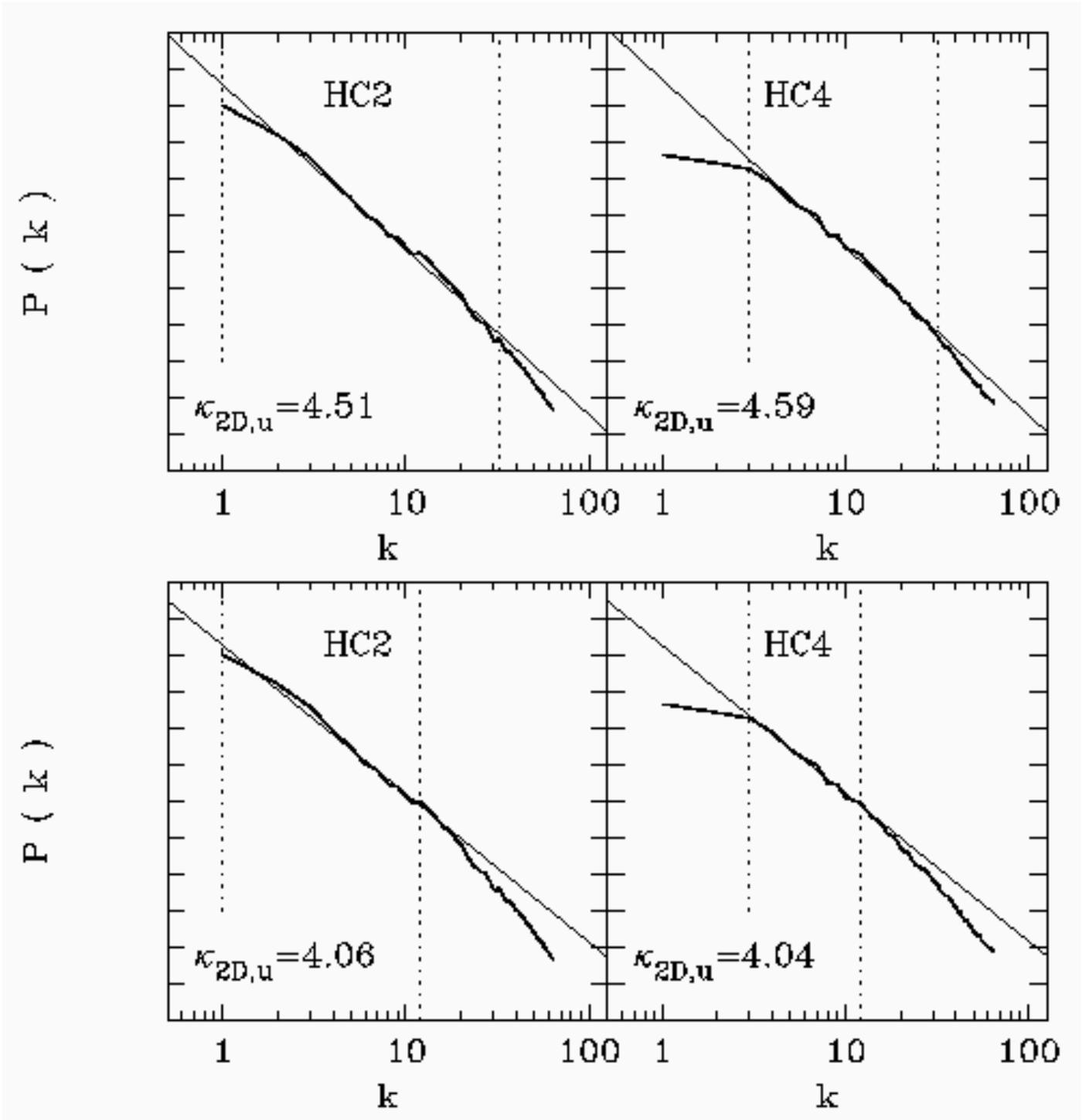}
\caption{Velocity line centroid power spectra for models HC2 and HC4. The dotted lines delimit the range of $k$ used in the power law fits (straight lines) to $\kappa_{2D,u}$.}
\label{fig:hcspecs}
\end{figure}

Example line centroid power spectra for the
HC2, HC4 models, driven at 
$k_{d}$~=~1--2 and 3--4  respectively,
are shown in Figure~\ref{fig:hcspecs}.
These are fitted over two ranges in $k$:
up to $k$~=~32 (upper panels) and up to
$k$~=~12 (lower panels); in both cases the 
smallest $k$ used to make the fit is taken 
as the smaller of the driving wavenumbers.
When the fit is
made to $k$~=~32, the spectra are ``steep''
($\kappa_{2D,u}$~$\sim$~4.5) but when the
fit is made to $k$~=~12, the indices are
in good agreement with the value expected
for Burgers turbulence ($\kappa_{2D,u}$~$\sim$~4).
Numerical dissipation will cause just the sort 
of steepening of the spectrum (less structure at 
small scale) that is seen in Figure~\ref{fig:hcspecs}.

\begin{figure}
\plotone{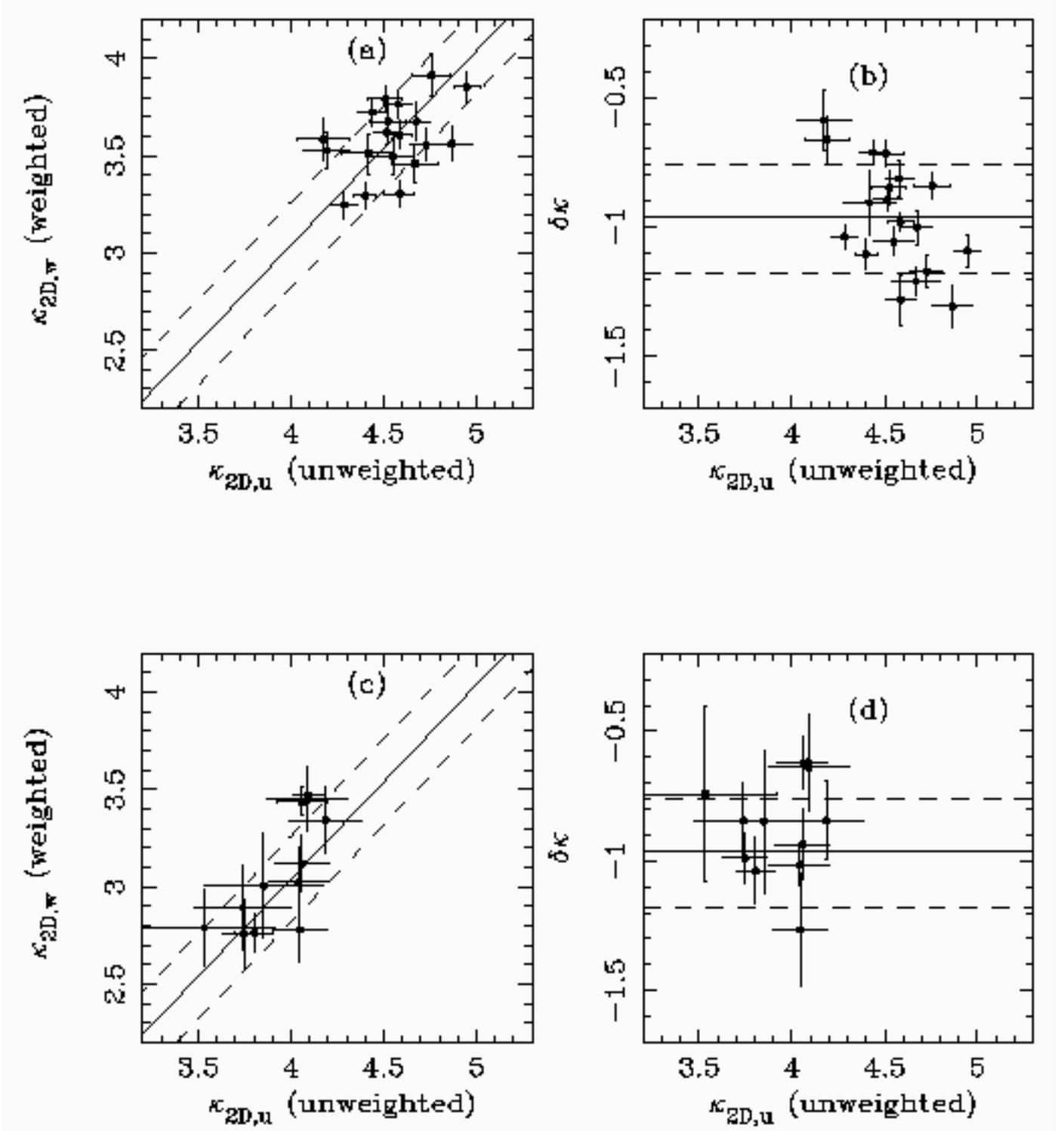}
\caption{(a) Plot of the density-weighted velocity line centroid spectral index, $\kappa_{2D,w}$, versus the unweighted (true) velocity line centroid spectral index, $\kappa_{2D,u}$ for all driven runs, with the indices fitted to $k$~=~32. (b) Plot of $\delta\kappa$ (=~$\kappa_{2D,w}$--$\kappa_{2D,u}$) versus $\kappa_{2D,u}$ for all driven runs, with the indices fitted to $k$~=~32. (c) Plot of the density-weighted velocity line centroid spectral index, $\kappa_{2D,w}$, versus the unweighted (true) velocity line centroid spectral index, $\kappa_{2D,u}$ for a subset of the driven runs (see text), with the indices fitted to $k$~=~12. (b) Plot of $\delta\kappa$ (=~$\kappa_{2D,w}$--$\kappa_{2D,u}$) versus $\kappa_{2D,u}$ for a subset of the driven runs (see text), with the indices fitted to $k$~=~12.  In all plots the solid line denotes the relationship $\delta\kappa$~=~--0.96 and the dashed lines mark uncertainties of $\pm$~0.22 around this relationship.}
\label{fig:absk}
\end{figure}

We have derived
$\kappa_{2D,u}$, $\kappa_{2D,w}$ and
$\delta\kappa$ for the driven runs over the two
ranges in $k$ given above. Figure~\ref{fig:absk}(a)
shows the relation between $\kappa_{2D,u}$
and $\kappa_{2D,w}$ for {\it all} driven runs fitted
to $k$~=~32;
the corresponding variation of $\delta\kappa$ with 
$\kappa_{2D,u}$ is shown in Figure~\ref{fig:absk}(b).
Figures~\ref{fig:absk}(c) and ~\ref{fig:absk}(d)
show the same relations obtained when the
fits are made to $k$~=~12 for the subset of runs with
$k_d$~=~1--2 and $k_d$~=~3--4. In all these Figures,
the solid line denotes $\delta\kappa$~=~--0.96
derived in the preceding Section, and the
dashed lines mark uncertainties of $\pm$0.22 
around this value.

These tests show that our estimated mean
$\delta\kappa$~=~--0.96~$\pm$~0.22 is not
strongly affected by the fitting range in $k$ but
there is some uncertainty in the absolute
range of $\kappa_{2D,u}$ and $\kappa_{2D,w}$
to which this value applies. Conservatively,
the fits to $k$~=~12 should be used, as
these are free of dissipative effects. Using
only the fields for which such fits can be
made, we find that $\delta\kappa$~=~--0.87~$\pm$~0.19,
and this applies to $\kappa_{2D,u}$~$\approx$~3.5--4.2
and $\kappa_{2D,w}$~$\approx$~2.7--3.4.

However, we also note that the lack of dependence of
$\delta\kappa$ on the fitting range in $k$
is indirect evidence that a single
value of $\delta\kappa$ is applicable at
all $\kappa_{2D,u}$, but this interpretation
should be examined in more detail in future.
In what follows we shall assume that 
$\delta\kappa$~=~--0.96~$\pm$~0.22 holds
for all $\kappa_{2D,u}$ and $\kappa_{2D,w}$.

\subsection{Origin of the Spectral Modification{\label{sec:origin}}}

\subsubsection{Effect of Density Spectral Index{\label{sec:kappacd}}}

Lazarian \& Esquivel (2003)
have suggested that (in our terminology) $\delta\kappa$
may have a larger amplitude for ``shallow'' density
field power spectra. 
The density field power spectrum
slope may be estimated from the power spectrum of
the projected column density field, $P_{cd}(k)$, using
the same arguments as presented in \S~\ref{sec:linecentroid}
(see also Lazarian \& Pogosyan 2000; Stutzki et al. 1998).
If $\delta\kappa$ is related to the spectral slope, $\kappa_{cd}$, of
the column density power spectrum (where $P_{cd}(k)$~$\propto$~$k^{-\kappa_{cd}}$)
then this would indeed provide a means of inferring
$\delta\kappa$ observationally. In other words, the
observed column density power spectrum slope is used as
a probe of the (relative) density fluctuation spectrum
as a function of scale within the medium. 

To investigate this, we produced column density
maps for all the simulations (including projection
along two axes -- parallel and perpendicular to the
magnetic field -- for the MHD fields). We then fitted
$\kappa_{cd}$ for each field, using the same range
in wavenumber that was used to derive $\delta\kappa$.
The fitted values of $\kappa_{cd}$ are given
in Table~\ref{tab:numericaldata}.
(The column density power spectra are also
slightly curved, so $\kappa_{cd}$ would be
lower if fitted to $k$~=~12; thus the $\kappa_{cd}$
axis is uncertain by a $\sim$linear shift as in
Figure~\ref{fig:absk} for $\kappa_{2D,u}$.)

\begin{figure}
\plotone{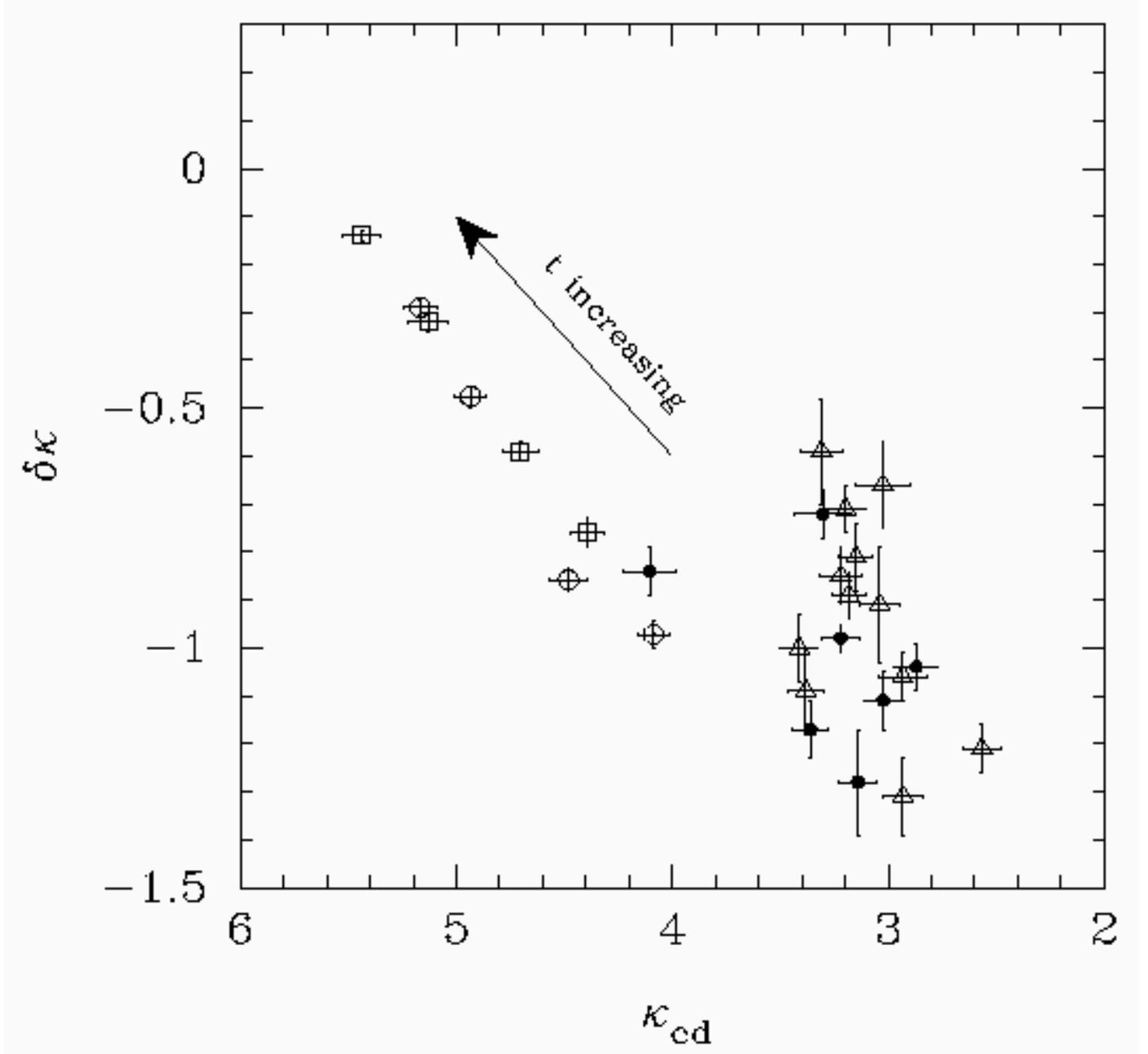}
\caption{Plot of $\delta\kappa$ versus $\kappa_{cd}$ for all simulations. Filled circles are driven HD; open triangles are driven MHD; open squares are decaying HD (model D, initial $M$~=~5); open circles are decaying HD (model U, initial $M$~=~50). The arrow shows the direction of evolution of $\delta\kappa$ as time (since driving is turned off) increases; see Figure~\ref{fig:dtime}.}
\label{fig:dkvskcd}
\end{figure}

In Figure~\ref{fig:dkvskcd} we plot the variation of
$\delta\kappa$ with $\kappa_{cd}$. There is a clear 
variation of $\delta\kappa$ with $\kappa_{cd}$, demonstrated
by the decaying turbulence simulations. As time proceeds, 
$\kappa_{cd}$ increases (i.e. preferential reduction
in small scale density contrast) and the amplitude of $\delta\kappa$
diminishes
(i.e. lessened effect of density inhomogeneity on the
velocity line centroid field). For the driven simulations,
the $\kappa_{cd}$ are typically $\sim$~3.
These are in reasonable accord with
observational estimates of $\kappa_{cd}$~$\approx$~2.8 for
molecular clouds; e.g. Bensch, Stutzki,
\& Ossenkopf 2001. For our $\kappa_{cd}$~$\sim$~3,
$\delta\kappa$ varies around the mean value of --1. 
Thus, observational estimates
of $\kappa_{cd}$ suggest that accounting for density inhomogeneity 
{\it is} very important for velocity line centroid analysis of molecular
clouds, implying (with $\delta\kappa$~$\approx$~--1) that
$\gamma_{2D}$~$\approx$~$\gamma_{3D}$ counter to the
expectations based on projection smoothing alone. This
provides an explanation for the $\gamma_{2D}$~$\approx$~$\gamma_{3D}$
result observed in these simulations by Ossenkopf \& Mac Low (2002).
Lazarian \& Pogosyan do not provide a 
physical motivation for whether the medium 
is density dominated (i.e. ``shallow density spectrum'' 
having $\kappa_{cd}$~$\leq$~3) or not: it is simply a 
question of what $\kappa_{cd}$ is, regardless of why.
In our scenario, this regime is the driven turbulence 
regime. 

\subsubsection{Amplitude of Density Fluctuations{\label{sec:sigmarho}}}

A note of caution regarding the use
of $\kappa_{cd}$ to infer $\delta\kappa$
should be made: $\kappa_{cd}$ is a non-unique
measure of the density field.
To investigate this more
deeply, we analyzed modified versions of
the HE2 field. First, we randomized the
Fourier space phases of the density and velocity fields
to produce new fields with the same power
spectra as the original fields. 
This modification has two
effects : (1) it removes the density-velocity
correlations, (2) it drastically reduces the
amplitude of density fluctuations.
Following phase-randomization of the
density field, we found that we had to
recenter the mean density due to the 
creation of negative densities
by the phase-randomization process. After
renormalization back to a mean density of
unity, the phase-randomized density field had
a standard deviation of 0.21 and range
of 1$\times$10$^{-6}$ to 2.02.
For comparison,
the original density field, normalized to
mean unity, had a standard deviation of
2.53 and a range of 1.2$\times$10$^{-6}$ to
92.1. The logarithm of the phase-randomized
density field had a standard deviation of
0.22, compared to the standard deviation of
the logarithm of the original density field
of 2.49.
Using the phase-randomized density fields and
velocity fields we 
derived $\delta\kappa$~=~--0.18~$\pm$~0.03.
This is substantially lower than the
original $\delta\kappa$~=~--1.04~$\pm$~0.05
obtained with the original HE2 density
and velocity fields. This shows that
the ``shallowness'' of the density
power spectrum, as measured by
$\kappa_{cd}$, does not uniquely 
allow inference of $\delta\kappa$.

We conducted a second test, this
time randomizing the phases of the
{\it logarithm} of density, and then exponentiating
the phase-randomized log(density) field.
This procedure again removes the density-velocity
correlations but does not lead to the
creation of negative densities, and allows
a greater amplitude of density contrast. 
We found that the resulting $\kappa_{cd}$~$\approx$~2.41
for the density field produced by this
method was slightly shallower than the original
$\kappa_{cd}$~$\approx$~2.87 for the HE2 model. 
The standard deviation of log(density) was
the same for the phase-randomized density field and
the original density field (2.49).
Using the exponentiated phase-randomized log(density)
field in conjunction with the phase-randomized velocity
field, we found $\delta\kappa$~=~--1.16~$\pm$~0.05.

These two tests together suggest that it is
the amplitude of density fluctuations, measured
as the standard deviation of log(density) that
is controlling the value of $\delta\kappa$.
To make sure that this is the case, and that
density-velocity correlation does not play a
major role, we took the original velocity
field and the original density field, shifted
by 64~pixels from its original position. For
a shift along the line-of-sight, we derived 
$\delta\kappa$~=~--0.90~$\pm$~0.04 and for a
shift transverse to the line-of-sight we
derived $\delta\kappa$~=~--0.97~$\pm$~0.06.
These are slightly lower in magnitude than the original 
$\delta\kappa$~=~--1.04~$\pm$~0.05, but the
difference is marginally significant.

\begin{figure}
\plotone{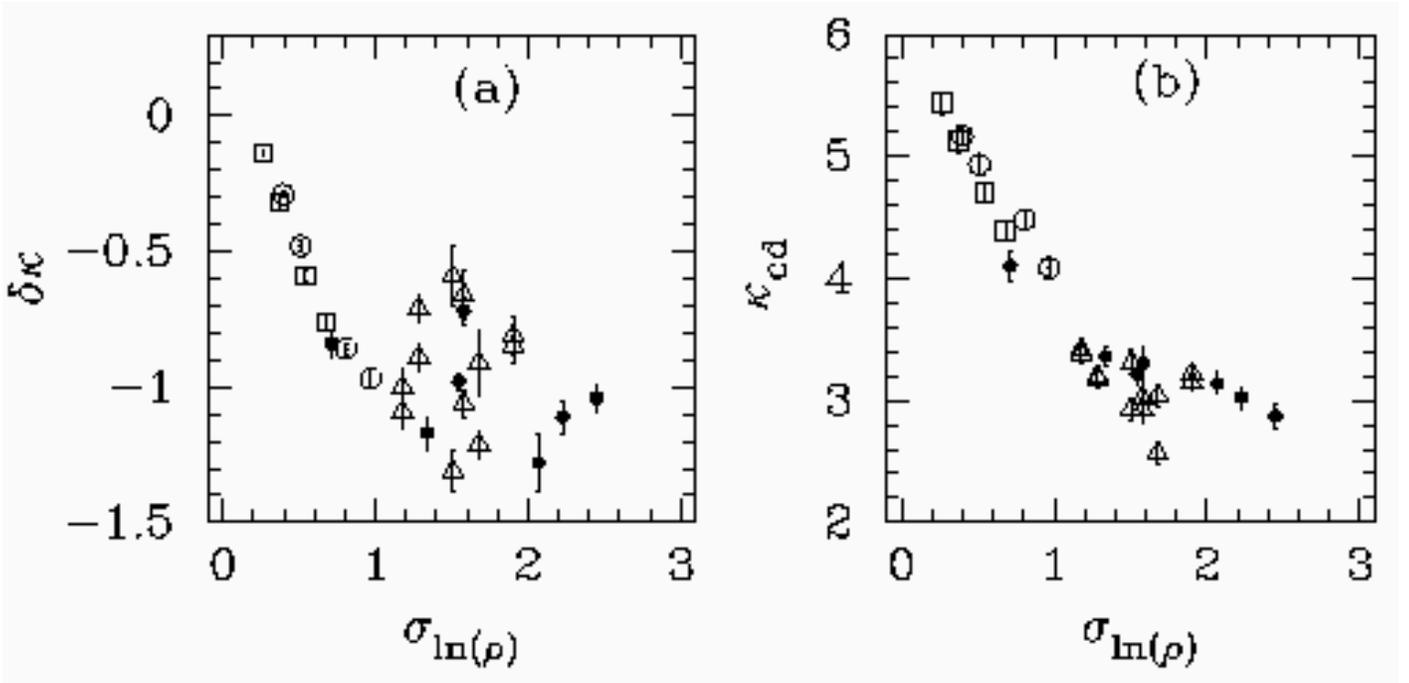}
\caption{(a) Plot of $\delta\kappa$ versus $\sigma_{ln(\rho)}$ for all simulations. (b) Plot of $\kappa_{cd}$ versus $\sigma_{ln(\rho)}$ for all simulations. Filled circles are driven HD; open triangles are driven MHD; open squares are decaying HD (model D, initial $M$~=~5); open circles are decaying HD (model U, initial $M$~=~50). The phase-randomized density model has $\sigma_{ln(\rho)}$~=~0.22, $\delta\kappa$~=~--0.18, $\kappa_{cd}$~=~2.87.}
\label{fig:siglogrhovsdk}
\end{figure}

We thus conclude that the amplitude of
density contrast is most important in 
determining $\delta\kappa$. To examine
this directly, we measured the
standard deviation of log(density), 
$\sigma_{ln(\rho)}$, for all the numerical
fields (see Table~\ref{tab:numericaldata})
and we plotted this versus 
$\delta\kappa$ (Figure~\ref{fig:siglogrhovsdk}(a))
and against $\kappa_{cd}$ Figure~\ref{fig:siglogrhovsdk}(b)).
This indicates that there is a fairly
good correlation between $\delta\kappa$
and $\sigma_{ln(\rho)}$, and that {\it with the
assumption} that the numerical fields are
a better representation of real molecular
clouds than the phase-randomized fields, that
$\kappa_{cd}$ may be used as a surrogate for
$\sigma_{ln(\rho)}$. We find that the
driven simulations are characterized
by $\kappa_{cd}$~$\approx$~3 and 
$\sigma_{ln(\rho)}$ exceeding $\sim$~unity,
but that there is no obvious dependence
of $\delta\kappa$ on $\kappa_{cd}$ or
$\sigma_{ln(\rho)}$ {\it within} this regime.

\begin{figure}
\plotone{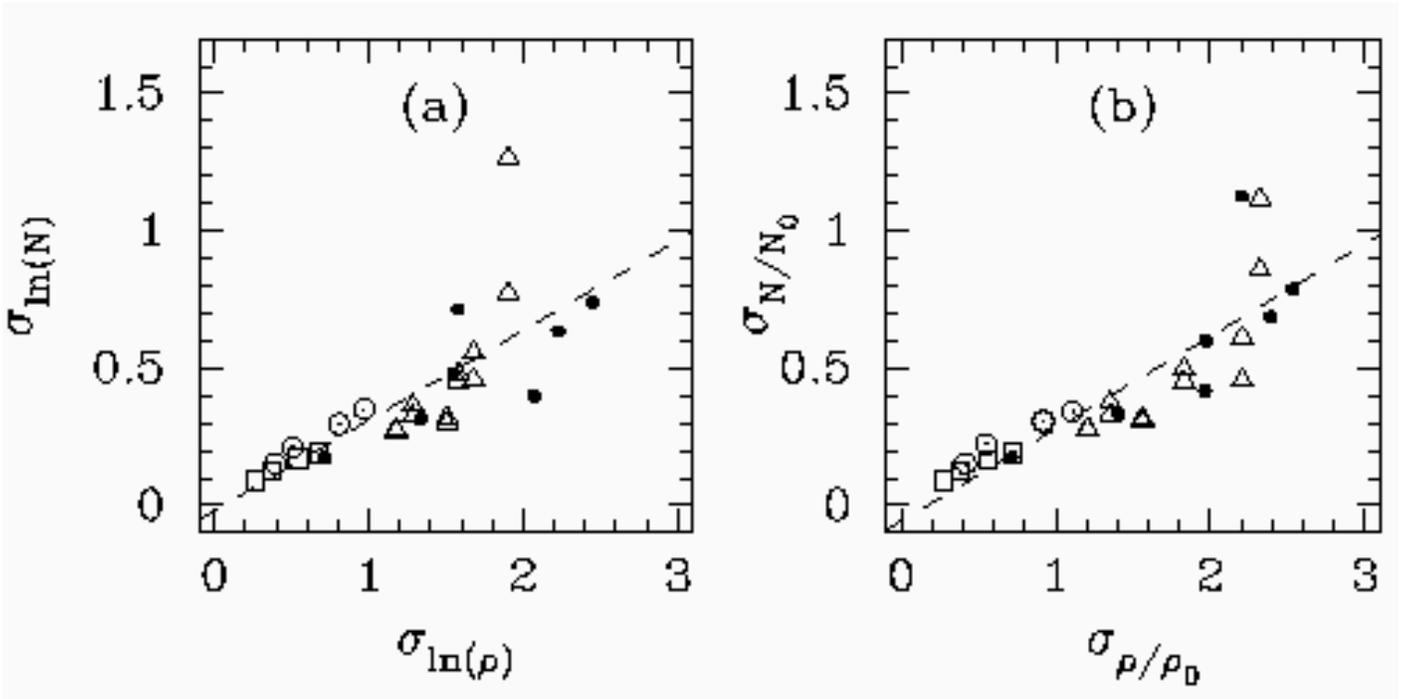}
\caption{(a)Plot of $\sigma_{ln(N)}$ versus $\sigma_{ln(\rho)}$ for all simulations. (b) Plot of $\sigma_{N/N_{0}}$ versus $\sigma_{\rho/\rho_{0}}$ for all simulations. Filled circles are driven HD; open triangles are driven MHD; open squares are decaying HD (model D, initial $M$~=~5); open circles are decaying HD (model U, initial $M$~=~50).}
\label{fig:Nversusrho}
\end{figure}

Ideally, one would like to measure
$\sigma_{ln(\rho)}$ to infer $\delta\kappa$
but it is of course impossible to measure
$\sigma_{ln(\rho)}$ directly. However,
it may be possible to infer $\sigma_{ln(\rho)}$
indirectly using the column density field.
Figure~\ref{fig:Nversusrho}(a) compares the variation
in the standard deviation of log(column density),
$\sigma_{ln(N)}$, to $\sigma_{ln(\rho)}$. The dashed
line is fitted : $\sigma_{ln(N)}$~=~--0.02~$\pm$~0.08~+~(0.33~$\pm$~0.06)$\sigma_{ln(\rho)}$ as a guide. The values of
$\sigma_{ln(N)}$ are listed in Table~\ref{tab:numericaldata}.
Unfortunately, it is also impossible to use
$\sigma_{ln(N)}$ observationally, as the
small $N$ are dominated by noise. Instead,
we consider simply the standard deviation
of density, $\sigma_{\rho/\rho_{0}}$, and column 
density $\sigma_{N/N_{0}}$, where both fields are
normalized their mean values, $\rho_{0}$ and $N_{0}$,
respectively. The dependence of $\sigma_{N/N_{0}}$
on $\sigma_{\rho/\rho_{0}}$ is
shown in Figure~\ref{fig:Nversusrho}(b) along with the fitted line,
$\sigma_{N/N_{0}}$~=~--0.06~$\pm$~0.07~+~(0.34~$\pm$~0.04)$\sigma_{\rho/\rho_{0}}$.
These results show that the variation in
column density may be used to infer the
variation in (3D) density, but the dynamic
range in $N$ is compressed due to line-of-sight
averaging ($\sigma_{N/N_{0}}$~$\approx$~$\sigma_{\rho/\rho_{0}}$/3).
However, in general,
it should be recognized 
that, at any finite
resolution, the mean density and column
density are
well-estimated, but their standard deviations
are continuously variable with resolution
in the respective dimensions. 
It may be possible to attempt
to construct a resolution-independent
measure to compare 2D and 3D standard
deviation but this is beyond the scope
of the current work. 

\begin{figure}
\plotone{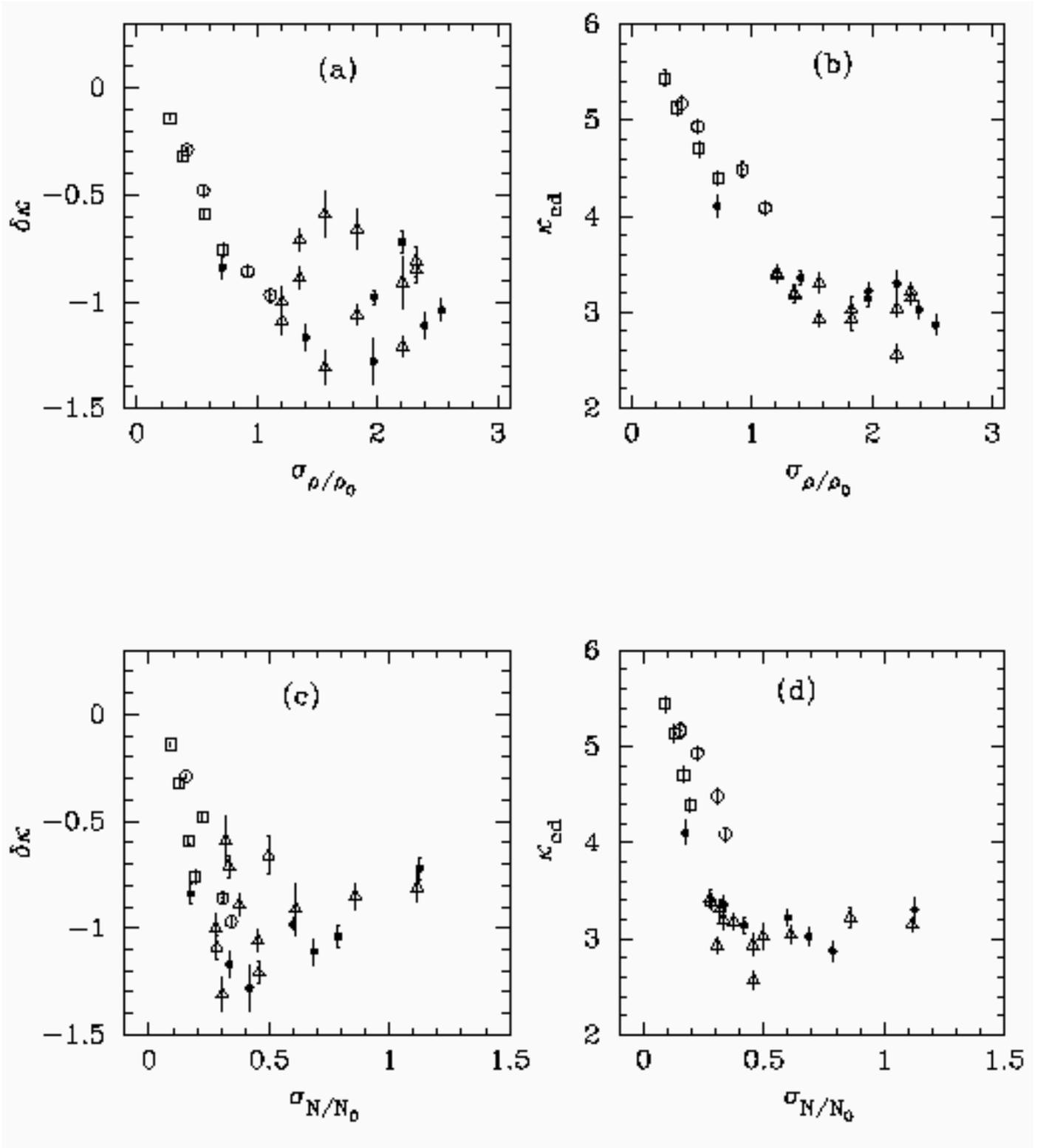}
\caption{(a) Plot of $\delta\kappa$ versus $\sigma_{\rho/\rho_{0}}$ for all simulations. (b) Plot of $\kappa_{cd}$ versus $\sigma_{\rho/\rho_{0}}$ for all simulations. (c) Plot of $\delta\kappa$ versus $\sigma_{N/N_{0}}$ for all simulations. (d) Plot of $\kappa_{cd}$ versus $\sigma_{N/N_{0}}$ for all simulations. Filled circles are driven HD; open triangles are driven MHD; open squares are decaying HD (model D, initial $M$~=~5); open circles are decaying HD (model U, initial $M$~=~50). The phase-randomized model has $\sigma_{\rho/\rho_{0}}$~=~0.21, $\sigma_{N/N_{0}}$~=~0.065, $\delta\kappa$~=~--0.18, and $\kappa_{cd}$~=~2.87. }
\label{fig:direct}
\end{figure}

In Figure~\ref{fig:direct}
we plot combinations of $\sigma_{\rho/\rho_{0}}$,
$\sigma_{N/N_{0}}$, $\delta\kappa$, and $\kappa_{cd}$
to gauge the utility of different measures.
In summary, some guidelines concerning
$\delta\kappa$ may be
established observationally
by measuring $\kappa_{cd}$
and $\sigma_{N/N_{0}}$. Note that
a measurement of $\kappa_{cd}$ and
$\sigma_{N/N_{0}}$ may also be used
to distinguish the phase-randomized fields
from the numerically simulated fields
: the phase-randomized density field has a much lower
$\sigma_{N/N_{0}}$~=~0.065 compared
with that of the numerical field ($\sigma_{N/N_{0}}$~=~0.786)
for the same $\kappa_{cd}$~=~2.87.
Some caution
comparing $\sigma_{N/N_{0}}$ with
Figure~\ref{fig:direct} is needed if
the observations are at significantly
different spatial dynamic range than
the simulations used here.

\subsection{Comparison with Previous Results{\label{sec:presults}}}

Our results appear inconsistent with two
previous studies. Brunt et al. (2003)
found that $\delta\kappa$~$\approx$~--0.4 from the
simulation used in their paper (Mach number $\approx$~1).
The density field used in Brunt et al. (2003)
had $\sigma_{ln(\rho)}$~=~0.45. Comparison 
with Figure~\ref{fig:siglogrhovsdk}(a) shows
that these numbers are not discrepant with
the current results. The conclusion of
Brunt et al. (2003) that, for a given amplitude
of density fluctuations, the correlations between
density and velocity are important was based
on the a single simulation and is outweighed
by our new results here.
Secondly, our results are inconsistent with the
results ($\delta\kappa$~$\approx$~0) found by 
Lazarian \& Esquivel (2003) for their
simulation (Mach number $\approx$~2.5). This
may be due to the solenoidal driving in their
simulations (Cho \& Lazarian 2002); certainly this
discrepancy deserves further attention.

\subsection{Resolution Tests{\label{sec:restests}}}

Finally, we examined the effect of limited
spatial dynamic range on our results. The driven
numerical fields implicitly contain realizations
at different effective resolutions (i.e. due to
the varying driving scales).
These show no obvious
trend of  $\delta\kappa$ with driving wavelength; however
there is a large amount of scatter in the derived
$\delta\kappa$ that may hide such a trend.
In addition we have explored the effect of limiting the
fits to $\delta\kappa$ to different ranges of $k$.
We conducted a further, more direct, tests using
realizations of model D (decaying turbulence)
at sizes of 128$^{3}$ and 64$^{3}$ and a realization of model HC2 (driven 
turbulence) at 256$^{3}$ resolution. These realizations
utilize the same {\it parameterization} but are
not identical {\it in detail} (i.e. they were not
initialized with the same random number seeds and so
are subject to the natural scatter in 
$\delta\kappa$ evidenced in Table~\ref{tab:numericaldata}).

Snapshots of model D at 128$^{3}$ and 64$^{3}$ sizes
are available at timesteps of 0.1, 0.15 and 0.3
sound crossing times. We conducted the same analysis
to derive $\delta\kappa$ from these lowered resolution
simulations. At 128$^{3}$ size, we find
$\delta\kappa$~=~--0.78~$\pm$~0.08, --0.57~$\pm$~0.04 , and
--0.22~$\pm$~0.03 for times 0.1, 0.15 and 0.3 respectively.
At 64$^{3}$ size, we find
$\delta\kappa$~=~--0.71~$\pm$~0.21, --0.33~$\pm$~0.22 , and
--0.11~$\pm$~0.25 for times 0.1, 0.15 and 0.3 respectively
As the resolution is increased, the $\delta\kappa$ are converging to
the values of $\delta\kappa$~$=$--0.76~$\pm$~0.03,--0.59~$\pm$~0.02, and
--0.32~$\pm$~0.02 found at 256$^{3}$ size
at the same timesteps.
The driven model HC2, with higher Mach number, showed a somewhat more 
pronounced shift, from $\delta\kappa$~$=$~--0.72~$\pm$~0.05 
at 128$^3$ to $\delta\kappa$~$=$~--1.01~$\pm$~0.03 at 256$^3$.  
In sum, this resolution study suggests that
the magnitude of $\delta\kappa$ is decreased slightly by
limited spatial dynamic range.  Therefore, our
derived $\delta\kappa$ magnitudes should probably be considered to
be lower limits (i.e. the true, high resolution, values
may be a little more negative than found here).

\section{Velocity Line Centroid Analysis : Observational Results{\label{sec:observationalresults}}}

\subsection{Overview{\label{sec:obsoverview}}}

\begin{figure}
\plotone{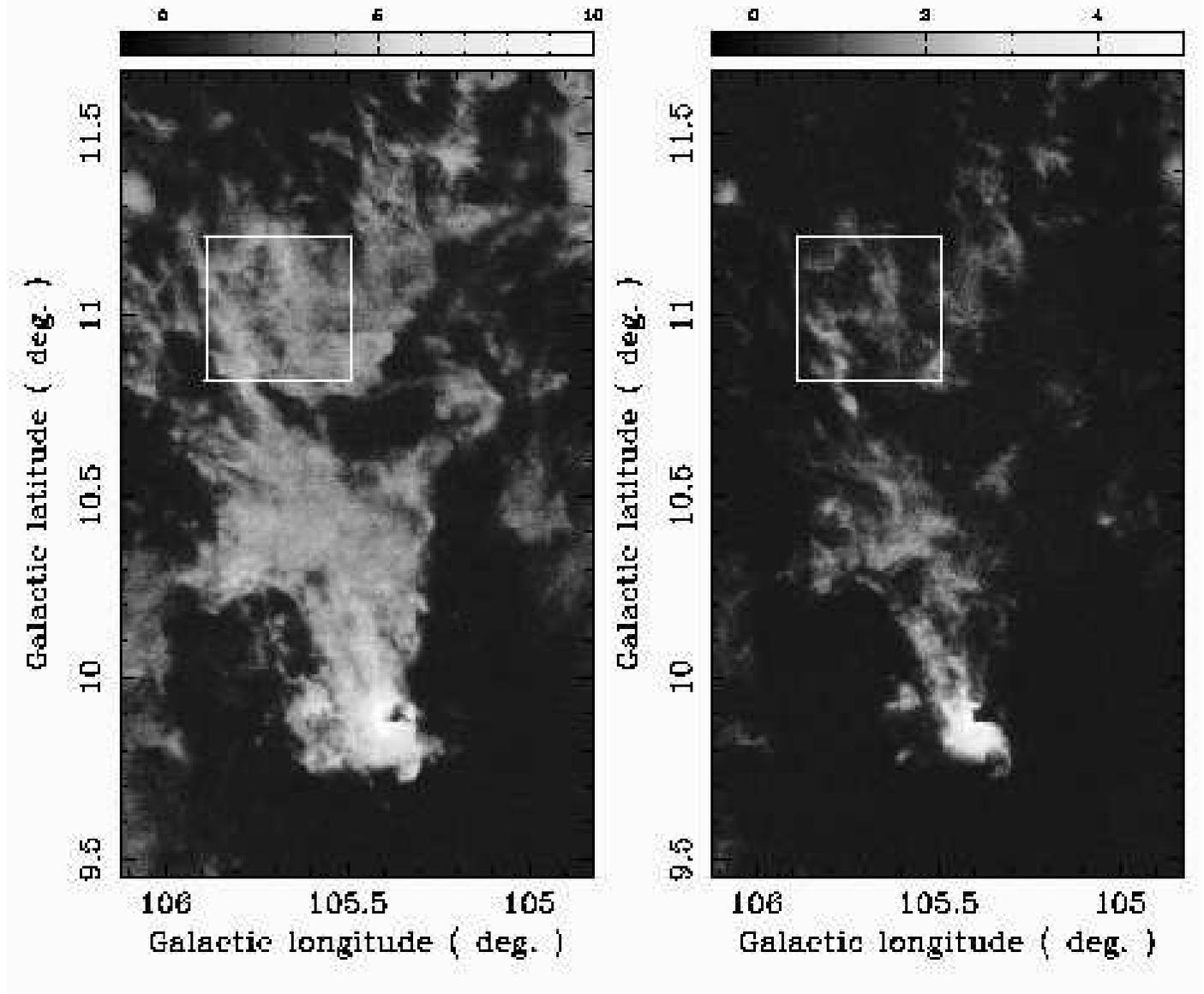}
\caption{Overview of the region near NGC~7129 selected for velocity line centroid analysis. Left : $^{12}$CO peak temperature image. Right : $^{13}$CO peak temperature image. The wedges show antenna temperature in K. }
\label{fig:ngc7129ima}
\end{figure}

\begin{figure}
\plottwo{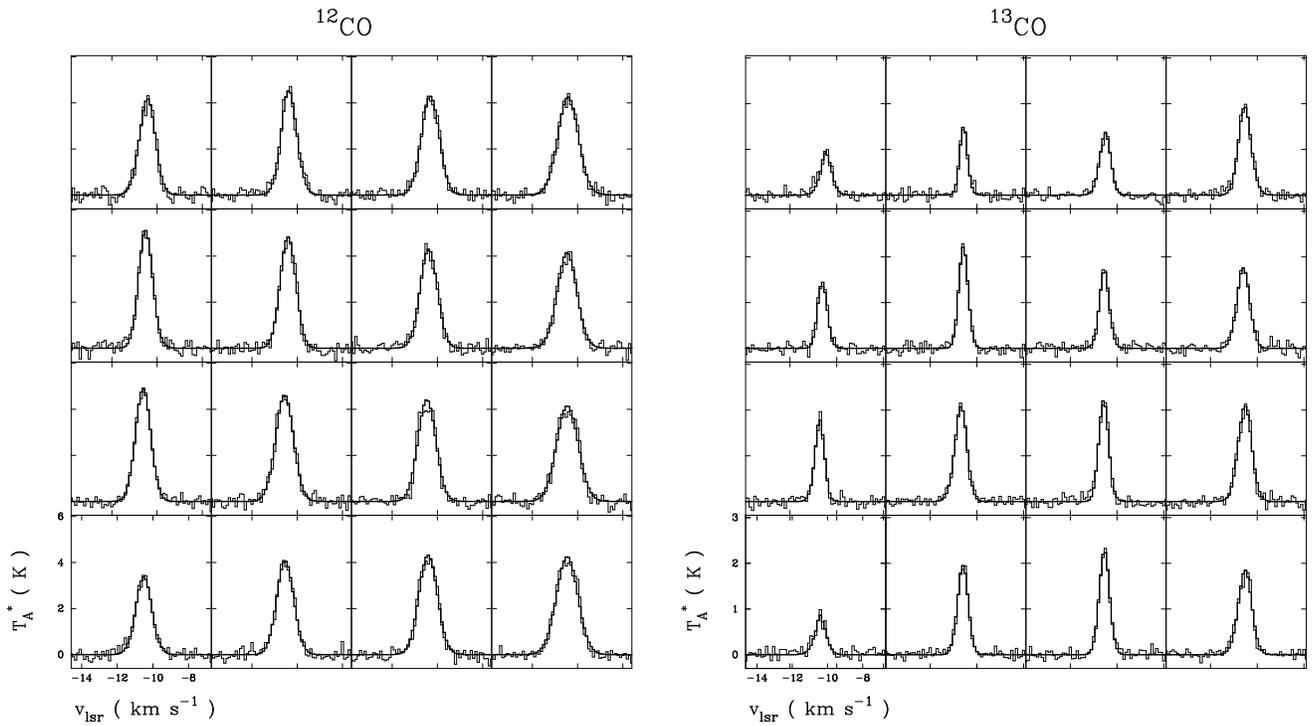}{f14b.eps}
\caption{Example fitted Gaussians (heavy lines) to the observed spectra (lighter lines) in the analysis region.}
\label{fig:spec12and13}
\end{figure}

An overview of the CGPS CO data in the vicinity
of NGC~7129 is given in Figure~\ref{fig:ngc7129ima}.
We have selected a region (64~$\times$~64 pixels, 
32~$\times$~32 beams, 4096 spectra) for velocity line centroid
analysis, indicated by the superimposed square
in Figure~\ref{fig:ngc7129ima}. This region is located
away from the main star forming area (it contains
no IRAS point sources), and has simple, singly-peaked
spectral lines. To derive velocity line centroid maps,
we utilized Gaussian fitting to the spectral lines rather
than a direct projection via Equation~4, as Gaussian
fitting was found to be less noisy (see below). 
Example fits to both the $^{12}$CO lines and $^{13}$CO
lines are shown in Figure~\ref{fig:spec12and13}. These
spectra are taken from a 16~$\times$~16 pixel block at
the center of the analysis region and are block-averaged over 
4~$\times$~4 pixels. The $^{12}$CO lines have roughly twice
the peak temperature of the $^{13}$CO lines, resulting in
comparable signal-to-noise for each tracer.

Fits to the $^{13}$CO lines could not be made for the
entirety of the analysis region, due to weak or absent
emission. Use of the $^{13}$CO line centroid field for
analysis would then require values for the missing data
to be assumed before calculation of the power spectrum.
A possible choice would be to set the missing data to
the mean value of the measurable data. However, a better
procedure is to create a ``hybrid'' line centroid field,
using the $^{12}$CO centroid velocities as the best
estimate of the missing $^{13}$CO centroid velocities.
We thus masked off all pixels in the analysis region
for which the peak temperature of the $^{13}$CO line
was less than 1.4~K $T_{A}^{*}$ (i.e. $\sim$~5$\sigma$),
and replaced the absent or noisy $^{13}$CO measurements with
the $^{12}$CO centroid velocities. For the hybrid field,
40\% of the line centroid velocities are from $^{13}$CO
and 60\% are from $^{12}$CO.

\begin{figure}
\plotone{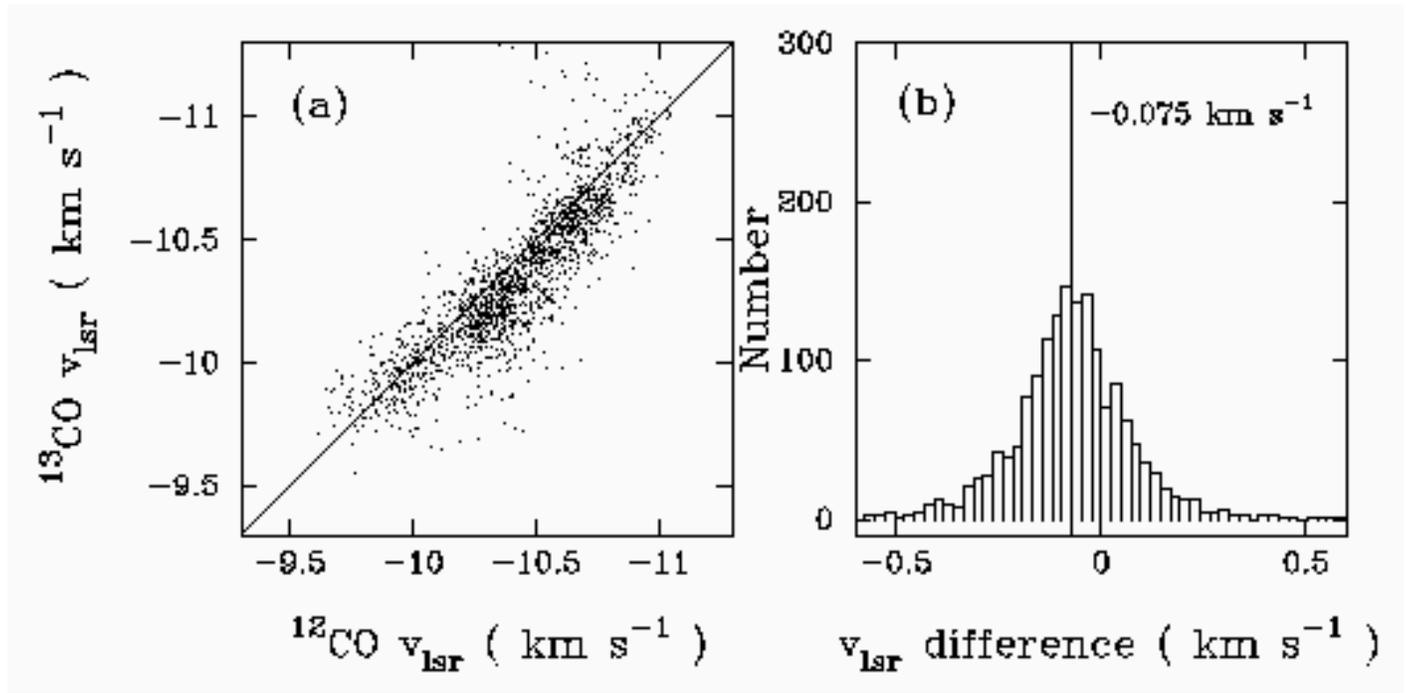}
\caption{(a) Comparison of the velocity line centroids derived from $^{12}$CO and $^{13}$CO where both measurements can be made. (b) Histogram of the differences in measured $v_{lsr}$ ($^{12}$CO velocity minus $^{13}$CO velocity).}
\label{fig:comp12and13}
\end{figure}

\begin{figure}
\plotone{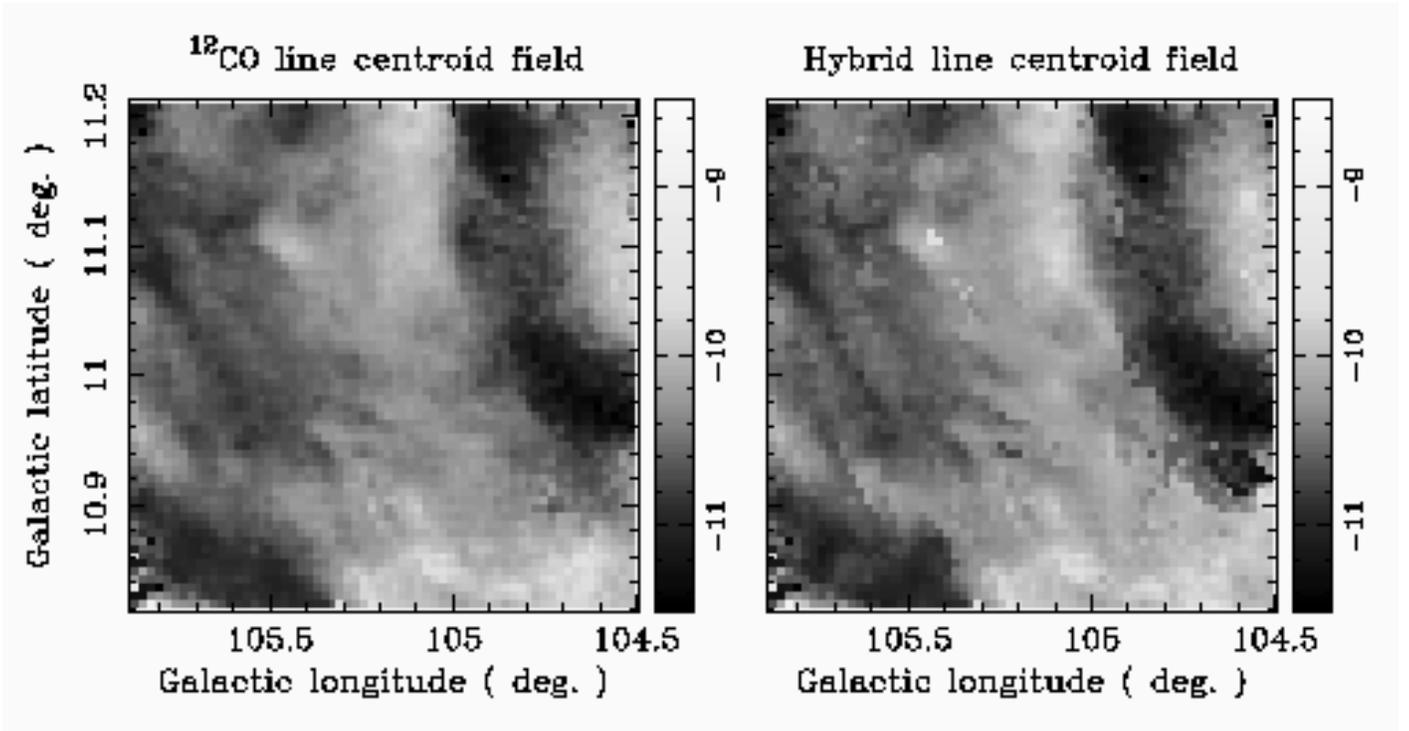}
\caption{The observed velocity line centroid maps.}
\label{fig:obslc}
\end{figure}

For the lines of sight where we have good fits
to both $^{13}$CO and $^{12}$CO, we compared the
derived line centroid velocities, as shown in
Figure~\ref{fig:comp12and13}(a). The two measurements
are very similar over the entire variation of
centroid velocity ($\sim$ --11 to --9.5~km~s$^{-1}$)
but there is a slight average offset between the two,
of --0.075~km~s$^{-1}$ ($\sim$~1/2 channel width) as
seen in Figure~\ref{fig:comp12and13}(b). This can arise
from greater saturation on one side of the $^{12}$CO line
profiles, or from depth-dependent abundance variations
in the presence of a line-of-sight velocity gradient.
The effects of opacity on our measured $^{12}$CO velocity line
centroid map are thus very small.
With the offset understood, the width of the 
distribution in Figure~\ref{fig:comp12and13}(b)
($\sim$~$\pm$~0.1~km~s$^{-1}$) can be otherwise attributed 
to noise on the measurements. To measure the noise
on the line centroid fields, we took the fitted Gaussian
line profiles as the best estimate of the true
line profiles and added actual instrumental noise from
the same spatial positions, taken from a signal-free
area of the baseline region. We then carried out the
same Gaussian fitting procedure to derive the
noise distribution. For comparison, we derived the
direct line centroid field from the ``re-noised'' fields
using Equation~4. The estimated errors on the $^{12}$CO
line centroid field are $\pm$~0.093~km~s$^{-1}$ (Gaussian
fitting) and $\pm$~0.122~km~s$^{-1}$ (using Equation~4).
The Gaussian-fitting
procedure is less noisy, and as the lines in the original
data are well-described by Gaussians (Figure~\ref{fig:spec12and13}),
we used the line centroids derived by Gaussian fitting for
the power spectrum analysis below. The resulting
velocity line centroid fields derived from Gaussian
fitting are shown in Figure~\ref{fig:obslc}. The observed fields
are qualitatively more similar to the simulations
driven at large scale 
(the $k_{d}$~=~1-2 case in Figure~\ref{fig:numlc}) than
to those driven at small scale.

\subsection{Large Scale Gradients and Edge Effects{\label{sec:edge}}}

Prior to the full analysis, we investigated the effect
of filtering out the ``large-scale'' line centroid 
velocity fluctuations. Whether this should be done or
not is a matter of some controversy (Miesch \& Bally 1994; Ossenkopf \& Mac Low 2002).
We estimated the large-scale gradients by fitting
a quadratic surface to the $^{12}$CO line centroid map.
Then we subtracted this from the original line centroid map,
and calculated the power spectrum before and after subtraction,
using FFTs.
We found that only the low frequency ($k$~=~1) power is strongly affected
by this procedure, as may have been expected. The measured
spectral slopes, between $k$~=~2~and~$k$~=~13 (see below)
are $\kappa_{2D}$~=~3.38~$\pm$~0.10 (before subtraction)
and $\kappa_{2D}$~=~3.42~$\pm$~0.09 (after subtraction).
Thus the spectrum steepens slightly (but insignificantly)
after subtraction of
the large-scale gradients, most likely due to the suppression
of high frequency structure resulting from mismatch in
the line centroids at the field edges (i.e. the FFT routine
assumes periodic fields). As there is little difference between
the line centroid power spectrum before and after subtraction
of the large-scale gradients, and it is not clear whether they
{\it should} be subtracted, we do not employ the subtraction
in the analysis below.

We also examined the effect of apodization (Miville-Desch\^enes,
Lagache \& Puget 2002). We centered the $^{12}$CO line
centroid field to zero mean, then applied a cosine taper
to a strip 2 pixels wide around the edge of the field.
The edges of the line centroid field now match at
zero velocity. For the apodized field, we
applied the same taper function to the noise estimates,
to ensure that the noise floor is still well estimated.
At large $k$ both the apodized line centroid power
spectrum {\it and} the apodized noise floor are reduced,
resulting in comparable signal-to-noise before and
after apodization. After subtraction of the
respective noise floors (see below) the fitted spectral slopes
are $\kappa_{2D}$~=~3.70~$\pm$~0.13 (before apodization)
and $\kappa_{2D}$~=~3.76~$\pm$~0.12 (after apodization).
There is little to distinguish these measurements
over the measurement uncertainty, so we do not utilize apodization
in the analysis below. As the apodization routine
simply imposes arbitrary smooth structure
in the line centroid field, it is not surprising that
the measured power spectrum slope steepens slightly
(i.e. reduction of small scale power). Alternatively,
the untapered field contains unphysical sharp velocity
structure at the field edges, so the actual noise-floor
subtracted spectral slope
most likely lies between the two slope estimates above.

\subsection{Correction for Noise and Beam Smearing{\label{sec:noiseandbeam}}}

The raw measured indices stated above ($\kappa_{2D}$~$\approx$~3.4)
are not the actual true measured indices, as we have not
yet accounted for the effects of measurement noise and
beam smearing on the measured power spectra. The total
power spectrum in the presence of noise is, simply,
$P_{true}(k)$~+~$P_{noise}(k)$ (where $P_{true}$ is the power spectrum
that would be observed in the absence of noise) assuming independence of
the signal and noise. The noise power spectrum must
thus be subtracted from the raw measured power spectrum
prior to analysis. The finite size of the telescope beam
induces correlations between closely spaced pixels; for
a beam pattern $B(x,y)$ in direct space (i.e. roughly a
Gaussian form with 45\arcsec~FWHM), the observed line
centroid map (neglecting noise for now) 
is : $v_{lc}^{obs}$~=~$v_{lc}^{obs,true}$~$\otimes$~$B$
where $\otimes$ denotes convolution and $v_{lc}^{obs,true}$
is the line centroid field that would be observed
without beam smearing. The beam smeared power spectrum
is then $P_{v_{lc}^{obs}}(k)$~=~$P_{v_{lc}^{obs,true}}(k)$$P_{B}(k)$
where $P_{B}(k)$ is the power spectrum of the beam
pattern. The final output power spectrum, corrected for
noise and beam smearing is:
$$ P_{output}(k) \;\;\; = \;\;\; (P_{v_{lc}^{obs}}(k) \; - \; P_{noise}(k)) \; / \; P_{B}(k). \;\; \eqno(5)$$
Note that the division out of the beam pattern is
applied {\it after} the noise floor subtraction as
the noises in our data are spatially independent (in contrast
to the measured signal).

\begin{figure}
\plottwo{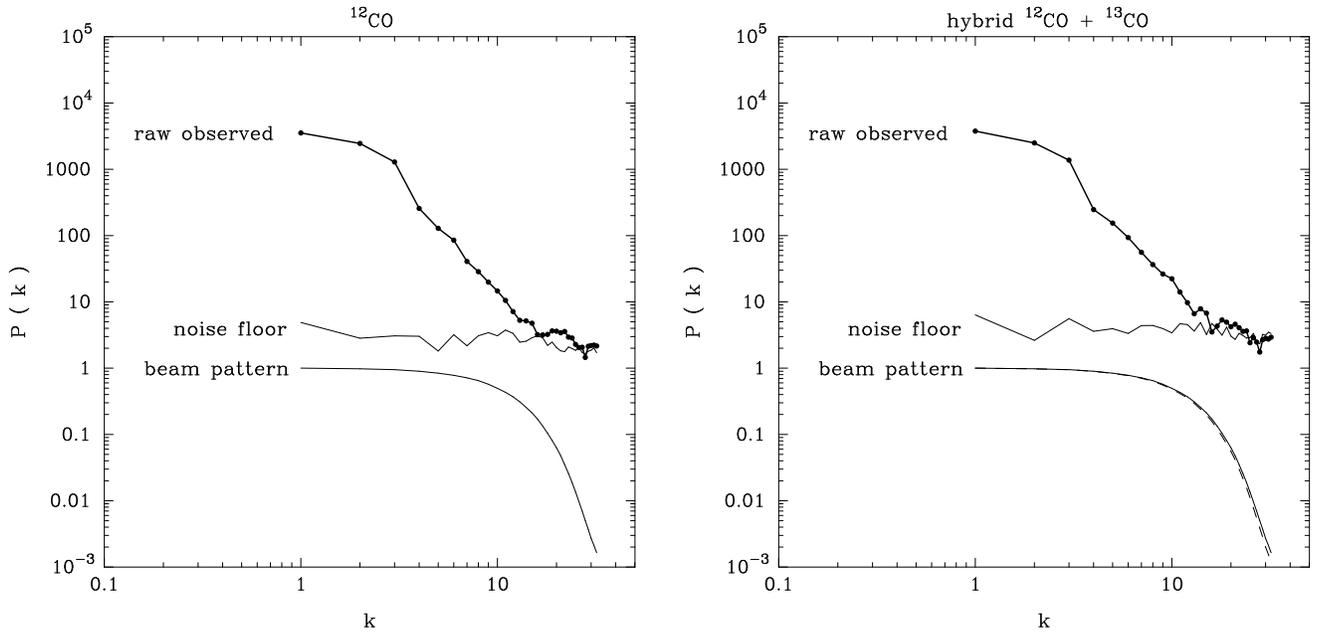}{f17b.eps}
\caption{Overview of the relevant terms necessary to construct the output line centroid power spectra. (a) $^{12}$CO field; raw observed power spectrum, noise floor and beam pattern (45\arcsec); (b) same for the hybrid field; the dashed line is the 46\arcsec~beam pattern applicable to the $^{13}$CO data; no uncertainty arises from applying the 45\arcsec~beam pattern to the hybrid field.}
\label{fig:rawspec}
\end{figure}

\begin{figure}
\plotone{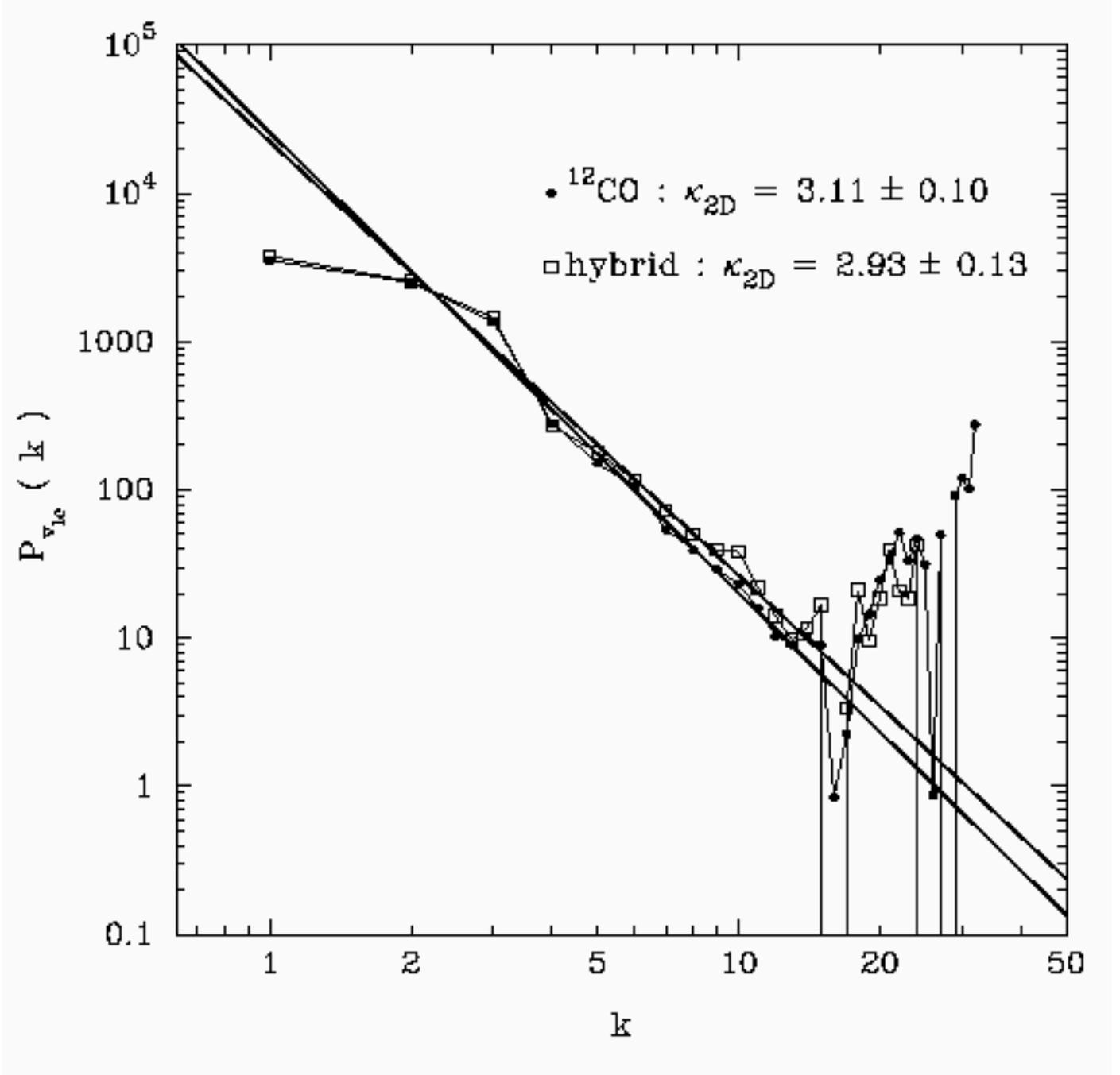}
\caption{Output velocity line centroid power spectra after noise floor subtraction and beam pattern division (Equation~5). The heavy lines are the fits of $\kappa_{2D}$ to these spectra in the range $k$~=~2 to $k$~=~13.}
\label{fig:finalspec}
\end{figure}

The noise power spectra can be derived from the ``re-noised'' fields
discussed above; we did this for both the $^{12}$CO line
centroid field and the hybrid line centroid field. We assumed
that a Gaussian beam pattern, $B(x,y)$ of 45\arcsec~FWHM applied
to both the $^{12}$CO line centroid field and the hybrid line centroid field.
A summary of the relevant terms in Equation~5 is provided in
Figure~\ref{fig:rawspec}.
Output power spectra were then derived via Equation~5 and
these are shown in Figure~\ref{fig:finalspec}. The 
resulting scaling range covers $k$~=~2 to $k$~=~13 limited
at low $k$ by a possible turnover and the
uncertainty of large-scale gradient subtraction, and
limited at high $k$ by the amplification of measurement
noise by beam pattern division. The resulting fits to
the range $k$~=~2 to $k$~=~13 are $\kappa_{2D}$~=~3.11~$\pm$~0.10
and $\kappa_{2D}$~=~2.93~$\pm$~0.13 for the $^{12}$CO line centroid field
and the hybrid line centroid field respectively. These are
jointly consistent with $\kappa_{2D}$~$\approx$~3,
giving $\gamma_{2D}$~$\approx$~0.5.

The range of the fit appears deceptively large in
Figure~\ref{fig:finalspec}; the spatial dynamic range
achieved is only 6.5, between wavelengths of 32 and
4.9 spatial pixels ($\sim$2.5 beamwidths). Similar
spatial dynamic range was achieved 
by Miesch \& Bally (1994), who measured a typical
value of $\gamma_{2D}$~$\approx$~0.43. Around three orders
of magnitude in spatial dynamic range was achieved
for the Polaris Flare molecular cloud
by Ossenkopf \& Mac Low (2002) by combining three data
sets at different angular resolutions and spatial scales;
they also measured $\gamma_{2D}$~$\approx$~0.5.

\subsection{Interpretation{\label{sec:obsinterp}}}

If we interpret the measured index of $\kappa_{2D}$~$\approx$~3
($\gamma_{2D}$~$\approx$~0.5) allowing for projection smoothing but
{\it without} taking into account the effect of
density weighting then the inferred 3D spectral 
index is $\kappa_{3D}$~$\approx$~3.
We would thus obtain 
$\gamma_{3D}$~$\approx$~0 (via Equation~2) or : no 
dependence of ``turbulent'' velocity
dispersion on scale. If correction for projection smoothing
is applied to the indices measured by Miesch \& Bally (1994)
and Ossenkopf \& Mac Low (2002), this again gives $\gamma_{3D}$~$\approx$~0.
A similar result (i.e. $\gamma_{3D}$~$\approx$~0) was found by
O'Dell \& Castaneda (1987) for \ion{H}{2} region turbulence,
after correcting for projection smoothing. The correction for
projection smoothing, however, has been the exception rather
than the rule for velocity line centroid analysis and most
authors have taken their projected $\gamma_{2D}$ as
being equivalent to the intrinsic $\gamma_{3D}$. This is
not correct, as the selection
of the $k_{z}$~=~0 plane of the 3D line of sight velocity 
power spectrum by projection onto the 2D plane of the
sky is purely a geometrical effect : projection smoothing
thus always applies (unless the cloud is actually two
dimensional -- see below), being modified only by the
effects of density inhomogeneity (and radiative transfer).
Hobson (1992) interprets (e.g.) $\kappa_{2D}$~=~3 (i.e. $\gamma_{2D}$~=~0.5)
as being applicable to shock-dominated turbulence ($\gamma_{3D}$~=~0.5).
Miesch \& Bally (1994) directly compare their $\gamma_{2D}$ to theoretical
predictions of $\gamma_{3D}$ with no modification. Lazarian \& Esquivel (2003)
derive from their numerical data (in our terminology)
$\delta\kappa$~=~0, but then interpret the
$\gamma_{2D}$~$\approx$~0.43 of Miesch \& Bally (1994) as
meaning also that $\gamma_{3D}$~$\approx$~0.43. 

Our numerical results here show
that the (statistical) equivalence of $\gamma_{2D}$ and $\gamma_{3D}$
is actually valid in the case of driven turbulence 
{\it but only due to the accidental
cancellation of two competing factors} : projection smoothing,
which suppresses small scale velocity line centroid structure, and
density inhomogeneity, which creates small scale 
velocity line centroid structure. 

\begin{figure}
\plotone{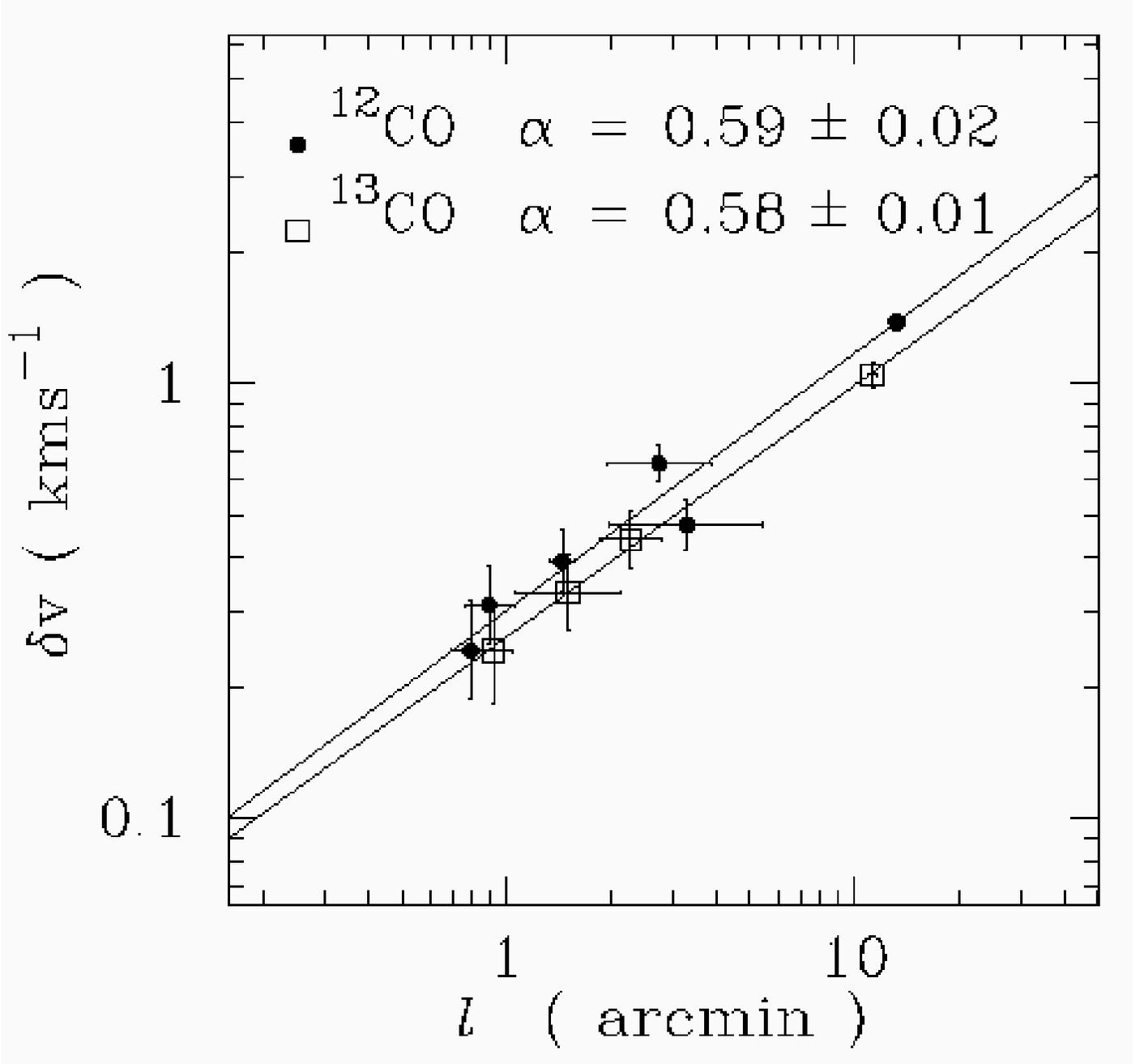}
\caption{Principal component analysis results for the selected field. The solid lines are the fits to the $^{12}$CO (upper) and $^{13}$CO (lower) ${\delta}v$, $l$ pairs (${\delta}v$~$\propto$~$l^{\alpha}$).}
\label{fig:pca}
\end{figure}

The interpretation of
existing results from line centroid analysis applied to
molecular clouds in general will depend on the inferred, or
assumed, state of turbulence.
If molecular clouds are in an evolved state
of decaying turbulence, then projection
smoothing {\it only} applies, and we 
would infer $\gamma_{3D}$~$\approx$~0 for
all clouds studied to date. 
If, however, the
clouds are still being strongly driven, or
have only recently formed, then the 
$\delta\kappa$~$\approx$~--1 induced by density 
inhomogeneity should apply, and we thus
can find a better consistency between line centroid
analysis and other measures of $\gamma_{3D}$.
To show this, we applied principal component
analysis (PCA) to the analysis region according
to the prescription given in Brunt \& Heyer (2002)
to derive a sequence of coupled characteristic 
scales, ${\delta}v$ and $\ell$, in velocity and space
respectively. 
The results of this, yielding slightly better than
an order of magnitude in spatial dynamic range,
are shown in Figure~\ref{fig:pca}. The ${\delta}v$ and $\ell$
pairs were fitted as ${\delta}v$~$\propto$~$\ell^{\alpha}$.
We derive $\alpha$~$=$~0.58~$\pm$~0.01 ($^{13}$CO)
and $\alpha$~$=$~0.59~$\pm$~0.02 ($^{12}$CO), 
resulting in $\gamma_{3D}$~$=$0.44~$\pm$~0.01 ($^{13}$CO)
and $\gamma_{3D}$~$=$0.46~$\pm$~0.04 ($^{12}$CO)
according to the Brunt (1999) calibration.
Some caution is necessary here, though, on two counts.
First, $\alpha$ is more directly dependent on
the first order velocity fluctuation spectrum, not
the second order spectrum that is traced by the
power spectrum of the line centroid field; second, 
estimates of $\alpha$ are uncertain (around the 
mean calibration line) by $\sim$0.1 (Brunt et al. 2003).
Neither of these factors can reconcile the PCA
results with $\gamma_{3D}$~$\approx$~0 however.
If, as suggested by PCA in the limit of weak
intermittency (as also supported by the lack of
strong exponential tails in the line profiles),
$\gamma_{3D}$~$\approx$~0.45 then we infer that
$\delta\kappa$~$\approx$~--0.9 would be needed to
reconcile the line centroid results with PCA.
This value of $\delta\kappa$ is consistent with
$\delta\kappa$~$=$~--0.96~$\pm$~0.22 derived from the 
numerical simulations of driven turbulence. 

\begin{figure}
\plotone{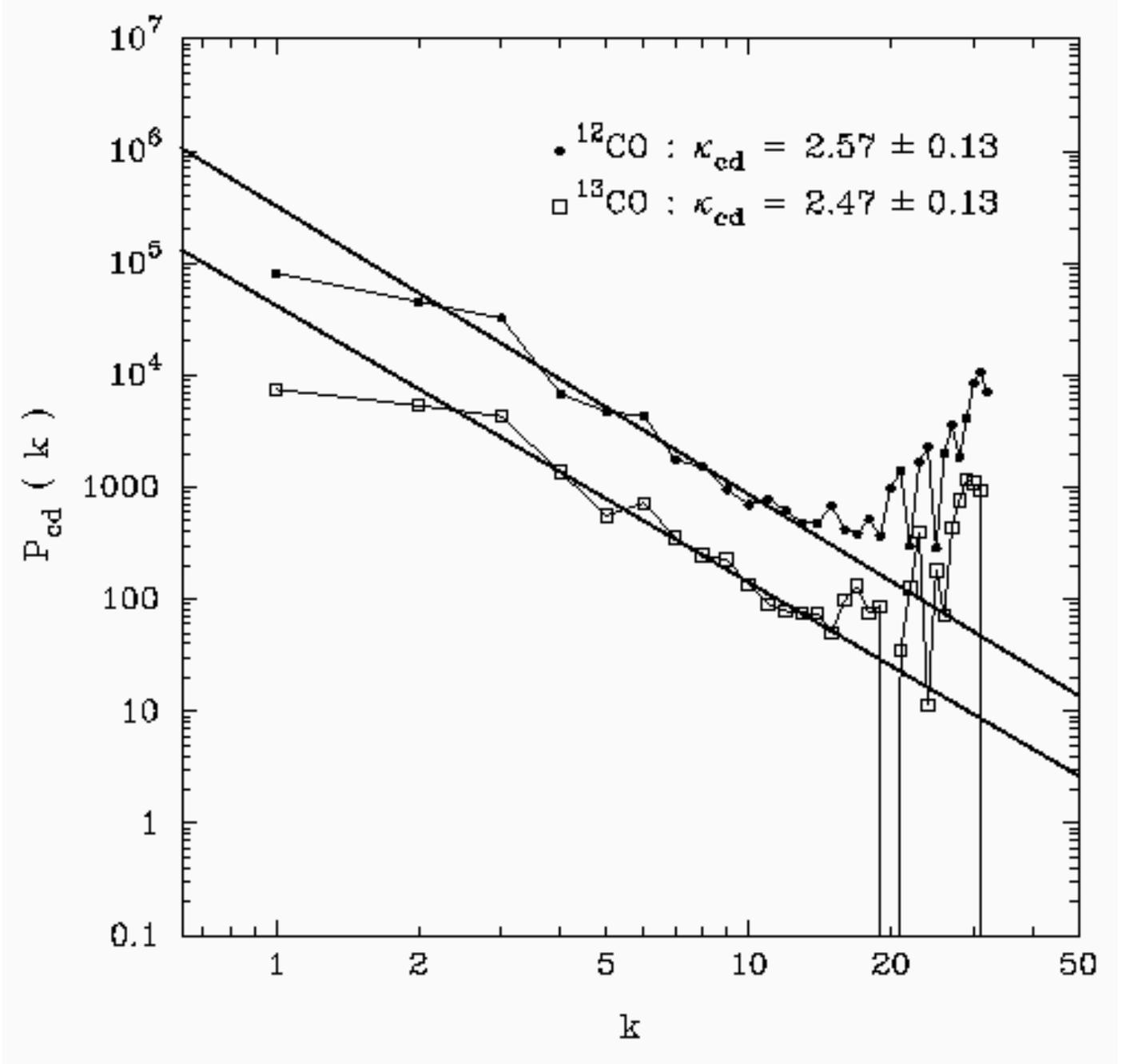}
\caption{Output integrated intensity power spectra after noise floor subtraction and beam pattern division. The heavy lines are the fits of $\kappa_{cd}$ to these spectra in the range $k$~=~2 to $k$~=~13.}
\label{fig:especcdobs}
\end{figure}

To determine the observationally-motivated
value of $\delta\kappa$, without recourse to PCA,
we examined the power spectra of the integrated intensity fields
for both isotopes (it would be very difficult to attempt
this with a hybrid integrated intensity field). Here,
we assume that the integrated intensity field provides
a good estimate of the column density field. To do this,
we estimated the noise floor by integrating signal-free
regions of the baseline over the same number of channels
as was used to generate integrated intensity maps. The
measured ``column density'' power spectra were then
corrected for the noise floor and beam pattern (Equation~5)
and the resulting output power spectra are 
shown in Figure~\ref{fig:especcdobs}.
We have fitted $\kappa_{cd}$ over the range $k$~=~2 to $k$~=~13 (the same
as used for the velocity line centroid power spectra), to estimate
$\kappa_{cd}$~$\approx$~2.5 consistently between the two
isotopes. This is in general agreement with typical
$\kappa_{cd}$ measured in molecular clouds (Bensch et al. 2001)
but slightly smaller than can be accounted for by our
numerical fields. Using Figure~\ref{fig:dkvskcd} as a rough guide, 
we interpret the measured $\kappa_{cd}$ as motivating 
$\delta\kappa$~$\approx$~--1. In doing this, we are
considering the numerically simulated fields to
be a more realistic representation of real molecular
clouds than the phase-randomized versions.
We do not see a turnover at small $k$ in
the integrated intensity power spectra, nor in the 
velocity line centroid power spectra, except perhaps at
$k$~=~1 which is quite uncertain. Our observed power spectra
are thus more in accord with the numerical models driven
at small wavenumber (see also Ossenkopf \& Mac Low 2002).
Finally, we used the $^{13}$CO data to provide
an estimate of $\sigma_{N/N_{0}}$, by assuming that
the $^{13}$CO is optically thin. To do this, we
created an integrated intensity image of the
emission and an integrated intensity image of
instrumental noise over the same number of
channels (64); the data were left in $T_{A}^{*}$ units,
since the absolute scaling is unimportant.
The mean integrated intensity of the data is :
$I_{0}$~=~1.034~$\pm$0.004~K~km~s$^{-1}$.
The total (signal+noise) standard deviation of
the integrated intensity image is $\sigma_{I,tot}$~=~0.661.
The contribution of the standard deviation of the noise is :
$\sigma_{I,noise}~=~$ 0.253~K~km~s$^{-1}$. Thus
we measure 
$\sigma_{I,signal/I_{0}}$~=~0.591~$\pm$0.004
which we take as being $\approx$~$\sigma_{N/N_{0}}$.
This is clearly inconsistent with 
$\sigma_{N/N_{0}}$~=~0.065 seen in the phase-randomized
density field (obtained with
$\kappa_{cd}$~=~2.87) and referral to
Figure~\ref{fig:direct}(d) will show a good
consistency between estimates of $\kappa_{cd}$ and
$\sigma_{N/N_{0}}$ for the observational
data compared to the numerical data.
The effective spatial dynamic range
in the simulations for computing $\sigma_{N/N_{0}}$
will lie between the nominal spatial dynamic
range (128 per axis) and the most stringent
interpretation of the maximum dissipation scale 
($\sim$ 12.8 per axis).
These compare reasonably well with the data, at a spatial 
dynamic range of 32 beams (64 pixels) per axis.
We interpret the consistency found between
$\kappa_{cd}$ and $\sigma_{N/N_{0}}$ as motivating
the use of the numerically simulated fields to
infer $\delta\kappa$~$\approx$~--1.

Our results regarding the effect of density
inhomogeneity indicate that the measured $\kappa_{2D}$~$\sim$~3
is thus {\it shallower} than the true $\kappa_{3D}$ by 
approximately $\delta\kappa$~$\approx$~--1, and we then
infer $\kappa_{3D}$~$\approx$~4 (c.f. Figure~\ref{fig:absk}(c))
and thus $\gamma_{3D}$~$\approx$~0.5. In more detail, we
derive (using $\delta\kappa$~=~--0.96~$\pm$~0.22)
$\kappa_{3D}$~=~4.07~$\pm$~0.24 ($^{12}$CO)
and $\kappa_{3D}$~=~3.89~$\pm$~0.26 (hybrid).
These convert to :
$\gamma_{3D}$~=~0.54~$\pm$~0.12 ($^{12}$CO)
and $\gamma_{3D}$~=~0.45~$\pm$~0.13 (hybrid).
These are in good agreement with $\gamma_{3D}$~$\approx$~0.45
determined from PCA, and are consistent
with compressible, Burgers turbulence. The
uncertainty in the inferred $\gamma_{3D}$ is large,
due mainly to the intrinsic 
variability of
$\delta\kappa$ ($\sim$~$\pm$~0.22).
We thus cannot entirely rule out the
Kolmogorov case $\kappa_{3D}$~$\approx$~3.67, 
$\gamma_{3D}$~$\approx$~0.33, from
velocity line centroid analysis alone, particularly
if we include the extra uncertainties arising from
subtraction of large scale gradients and edge
effects (estimated to be $\sim$~$\pm$~0.1 in $\kappa_{2D}$).

We interpret the conditions required for consistency between
velocity line centroid analysis, principal component analysis and
size-linewidth analysis
as providing indirect support for molecular
clouds that have only recently formed and/or are continually 
driven on large scales
(Elmegreen 1993; Ballesteros-Paredes, V{\'a}zquez-Semadeni, \& Scalo 1999). 

Projection smoothing does not apply in the case that the line of
sight depth over which the observational tracer exists is much less than the
total transverse scale of emission observed on the sky, requiring
sheet-like clouds, oriented face-on towards us. For a face-on sheet,
the effect of density weighting would also vanish as there is
no longer any line of sight variability (see Equation~3 in the limit that
$N_{pix}$~$\longrightarrow$~1 as a representation of this regime). 
While this solution is quite unsatisfactory in general, it 
represents a limiting case : no projection
smoothing, no density weighting effects, resulting in the direct measurement
of intrinsic statistical properties; i.e. the field 
is actually two dimensional. 
Some modified version of this limiting case may be operating in
our numerical fields; i.e. in the case that a single density enhancement (not
necessarily confined to a thin sheet, but having some spatial coherence) 
is dominant along the line of sight, then 
this will approximate the $N_{pix}$~$\longrightarrow$~1 limit of Equation~3.

In general, for the full 3D (isotropic) driven case, our numerical results
show that projection smoothing is countered by the effects of
density inhomogeneity, again resulting in recovery of intrinsic statistical
scaling properties, but in a less direct way.

\section{Summary{\label{sec:summary}}}

We have examined numerical simulations of supersonic turbulence
in molecular clouds in order to understand the effects of
projection smoothing and density inhomogeneity on observable
velocity line centroid fields ($v_{lc}(x,y)$). We find that the observed 
index, $\gamma_{2D}$, describing the variation of projected
line centroid velocities ($\langle {\delta}v_{lc}^{2} \rangle^{1/2}$~$\propto$~$\ell^{\gamma_{2D}}$)
is related to the intrinsic index, $\gamma_{3D}$, 
(where $\langle {\delta}v^{2} \rangle^{1/2}$~$\propto$~$\ell^{\gamma_{3D}}$) via
the following relation :
$$ \gamma_{2D} \;\;\; = \;\;\; \gamma_{3D} \;\;\; + \;\;\; 0.5 \;\;\; + \;\;\; \delta\kappa/2,$$
where the increase of 0.5 is a direct consequence of projection smoothing, and
$\delta\kappa$ is an empirically determined correction for the effects of
density inhomogeneity.

For driven turbulence, both with and without magnetic fields,
$\delta\kappa$~$\approx$~--1 (with variations $\sim$~$\pm$~0.22), 
thus implying $\gamma_{2D}$~$\approx$~$\gamma_{3D}$ due 
to the accidental
cancellation of the effects of projection smoothing
and small-scale structure induced by density 
inhomogeneity on the line centroid field. 
For decaying turbulence, $\delta\kappa$ tends to $\sim$0 at late
times, so that $\gamma_{2D}$~$\approx$~$\gamma_{3D}$~+~0.5 due
entirely to projection smoothing. 

The variations in $\delta\kappa$ observed in high Mach number driven
turbulence have no obvious dependence on Mach number, $M$.
As $M\longrightarrow$0 (the incompressible case)
one should also expect that $\delta\kappa\longrightarrow$0, as 
is indeed seen at late times in the decaying runs. We have
inferred that $\delta\kappa$ is determined mostly by
the amplitude of density fluctuations.
The scatter in $\delta\kappa$
in the supersonic regime emphasizes the need for continued
high spatial dynamic range surveying of the molecular ISM 
in order to provide a statistically useful sample of fields. 

Deprojection of observed scaling relations in velocity line
centroid fields requires some knowledge of the state of
the flow. This information can be inferred
from the spectral slope of the column density power spectrum
and a measure of the standard deviation in column
density relative to the mean ($\sigma_{N/N_{0}}$).
Our numerical results show that $\gamma_{3D}$ is now 
consistently estimated at $\sim$0.5 using
velocity line centroid analysis and using principal component 
analysis and size-linewidth relations.
This consistency requires that 
molecular clouds are continually driven or only recently formed.

\acknowledgements

We acknowledge useful comments from
Volker Ossenkopf and particularly
the anonymous referee.
The Dominion Radio Astrophysical
Observatory is a National Facility
operated by the National Research
Council. The Canadian Galactic Plane
Survey is a Canadian project with
international partners, and is supported
by the Natural Sciences and Engineering
Research Council (NSERC). CB is supported
by a grant from NSERC to the Canadian
Galactic Plane Survey. 
M-MML is supported by NSF CAREER grant 
AST99-85392 and NASA Astrophysical Theory 
Program grant NAG5-10103.  Computations 
analyzed here were performed at the Rechenzentrum 
Garching of the Max-Planck-Gesellschaft.
The Five College Radio Astronomy Observatory 
is supported by NSF grant AST 01-00793.



\begin{deluxetable}{ccccccccccc}
\tablewidth{0pt}
\tablecolumns{11}
\tablecaption{Numerical Models : Parameters and Measurements}
\tablehead{\colhead{Model} & \colhead{$N_{pix}$\tablenotemark{a}} & \colhead{HD/MHD/Decay} & \colhead{$k_{d}$\tablenotemark{b}} & \colhead{{\it M}\tablenotemark{c}} & \colhead{$v_{A}/c_{s}$\tablenotemark{d}} & \colhead{t\tablenotemark{e}} & \colhead{$\delta\kappa$} & \colhead{$\kappa_{cd}$} & \colhead{$\sigma_{ln(\rho)}$} & \colhead{$\sigma_{ln(N)}$}}
\startdata
HA8  &             128 &    HD    & 7--8 &  1.9  &  0 &     1  & --0.84 $\pm$ 0.05 & 4.10 $\pm$ 0.12 & 0.71 & 0.17  \\
HC2  &             128 &    HD    & 1--2 &  7.4  &  0 &     1  & --0.72 $\pm$ 0.05 & 3.30 $\pm$ 0.13 & 1.58 & 0.71  \\
HC4  &             128 &    HD    & 3--4 &  5.3  &  0 &     1  & --0.98 $\pm$ 0.03 & 3.22 $\pm$ 0.09 & 1.55 & 0.48 \\
HC8  &             128 &    HD    & 7--8 &  4.1  &  0 &     1  & --1.17 $\pm$ 0.06 & 3.36 $\pm$ 0.08 & 1.34 & 0.32  \\
HE2  &             128 &    HD    & 1--2 & 15.0  &  0 & 0.475  & --1.04 $\pm$ 0.05 & 2.87 $\pm$ 0.10 & 2.45 & 0.74 \\
HE4  &             128 &    HD    & 3--4 & 12.0  &  0 & 0.875  & --1.11 $\pm$ 0.06 & 3.02 $\pm$ 0.09 & 2.23 & 0.63  \\
HE8  &             128 &    HD    & 7--8 &  8.7  &  0 &     1  & --1.28 $\pm$ 0.11 & 3.14 $\pm$ 0.09 & 2.07 & 0.40  \\  
MC4X $v_{\perp}$ & 128 &   MHD    & 3--4 &  5.3  & 10 & 0.1  & --0.91 $\pm$ 0.12 & 3.04 $\pm$ 0.09 & 1.68 & 0.56   \\
MC4X $v_{||}$    & 128 &   MHD    & 3--4 &  5.3  & 10 & 0.1  & --1.21 $\pm$ 0.05 & 2.57 $\pm$ 0.09 & 1.68 & 0.46   \\
MC45 $v_{\perp}$ & 128 &   MHD    & 3--4 &  4.8  &  5 & 0.2  & --0.66 $\pm$ 0.09 & 3.03 $\pm$ 0.13 & 1.58 & 0.45    \\
MC45 $v_{||}$    & 128 &   MHD    & 3--4 &  4.8  &  5 & 0.2  & --1.06 $\pm$ 0.05 & 2.93 $\pm$ 0.11 & 1.58 & 0.48    \\
MC41 $v_{\perp}$ & 128 &   MHD    & 3--4 &  4.7  &  1 & 0.5  & --0.89 $\pm$ 0.05 & 3.18 $\pm$ 0.08 & 1.28 & 0.36   \\
MC41 $v_{||}$    & 128 &   MHD    & 3--4 &  4.7  &  1 & 0.5  & --0.71 $\pm$ 0.05 & 3.20 $\pm$ 0.09 & 1.28 & 0.32     \\
MC85 $v_{\perp}$ & 128 &   MHD    & 7--8 &  3.4  &  5 & 0.075  & --0.59 $\pm$ 0.11 & 3.31 $\pm$ 0.10 & 1.50 & 0.32  \\
MC85 $v_{||}$    & 128 &   MHD    & 7--8 &  3.4  &  5 & 0.075  & --1.31 $\pm$ 0.08 & 2.93 $\pm$ 0.09 & 1.50 & 0.30 \\
MC81 $v_{\perp}$ & 128 &   MHD    & 7--8 &  3.5  &  1 & 0.225  & --1.00 $\pm$ 0.07 & 3.41 $\pm$ 0.09 & 1.18 & 0.27  \\
MC81 $v_{||}$    & 128 &   MHD    & 7--8 &  3.5  &  1 & 0.225  & --1.09 $\pm$ 0.06 & 3.38 $\pm$ 0.08 & 1.18 & 0.28  \\
ME21 $v_{\perp}$ & 128 &   MHD    & 1--2 & 14.0  &  1 & 0.5    & --0.85 $\pm$ 0.06 & 3.22 $\pm$ 0.09 & 1.90 & 0.77   \\
ME21 $v_{||}$    & 128 &   MHD    & 1--2 & 14.0  &  1 & 0.5    & --0.81 $\pm$ 0.07 & 3.15 $\pm$ 0.08 & 1.90 & 1.27   \\
D                & 256 & HD/Decay & 1--8 &  5.0  &  0 & 0.1   & --0.76 $\pm$ 0.03 & 4.39 $\pm$ 0.08 & 0.68 & 0.19     \\
D                & 256 & HD/Decay & 1--8 &  5.0  &  0 & 0.15  & --0.59 $\pm$ 0.02 & 4.70 $\pm$ 0.08 & 0.54 & 0.17  \\
D                & 256 & HD/Decay & 1--8 &  5.0  &  0 & 0.3  & --0.32 $\pm$ 0.02 & 5.13 $\pm$ 0.09 & 0.37 & 0.13   \\
D                & 256 & HD/Decay & 1--8 &  5.0  &  0 & 0.5  & --0.14 $\pm$ 0.01 & 5.44 $\pm$ 0.09 & 0.26 & 0.09   \\
U                & 256 & HD/Decay & 1--8 & 50.0  &  0 & 0.1  & --0.97 $\pm$ 0.03 & 4.09 $\pm$ 0.07 & 0.97 & 0.35   \\
U                & 256 & HD/Decay & 1--8 & 50.0  &  0 & 0.15  & --0.86 $\pm$ 0.02 & 4.48 $\pm$ 0.09 & 0.81 & 0.30    \\
U                & 256 & HD/Decay & 1--8 & 50.0  &  0 & 0.3  & --0.48 $\pm$ 0.02 & 4.93 $\pm$ 0.07 & 0.51 & 0.21    \\
U                & 256 & HD/Decay & 1--8 & 50.0  &  0 & 0.5 & --0.29 $\pm$ 0.01 & 5.17 $\pm$ 0.08 & 0.40 & 0.15    \\
\enddata
\tablenotetext{a}{Number of pixels in each dimension}
\tablenotetext{b}{Driving wavenumber}
\tablenotetext{c}{rms Mach number (equilibrium $M$ for driven turbulence; initial $M$ for decaying turbulence)}
\tablenotetext{d}{Ratio of Alfv{\'e}n speed to sound speed}
\tablenotetext{e}{Time step, measured in sound crossing times}
\label{tab:numericaldata}
\end{deluxetable}

\clearpage

\clearpage

\end{document}